\documentclass[12pt]{article}
\usepackage{epsfig,amsfonts,amssymb}
\usepackage{hyperref}
\usepackage{comment}
\usepackage{cite}
\input epsf.sty
\usepackage{cite}

\def\ZZZ{{\hbox{ Z\kern-1.6mm Z}}}
\def\RRR{{\hbox{ R\kern-2.4mm R}}}
\def\CCC{{\hbox{ C\kern-2.0mm C}}}
\def\zzz{{\hbox{z\kern-1mm z}}}

\newcommand{\qeq}{{\hbox{=\kern-2.3mm ? \kern.5mm }}}
\renewcommand{\qeq}{=}

\newcommand{\eps}{\epsilon}

\newcommand{\vp}{\varphi}
\newcommand{\ve}{\varepsilon}

\newcommand{\bJ}{{\bf J}}

\newcommand{\OO}{{\cal O}}

\newcommand{\wt}{\widetilde}

\newcommand{\NN}{{\cal N}}

\newcommand{\be}{\begin{equation}}
\newcommand{\ee}{\end{equation}}
\newcommand{\ben}{\begin{eqnarray}\displaystyle}
\newcommand{\een}{\end{eqnarray}}

\newcommand{\refb}[1]{(\ref{#1})}
\newcommand{\p}{\partial}
\newcommand{\sectiono}[1]{\section{#1}\setcounter{equation}{0}}

\newcommand{\ia}{i}
\newcommand{\ja}{j}

\def\one{{\hbox{ 1\kern-.8mm l}}}
\def\zero{{\hbox{ 0\kern-1.5mm 0}}}

\newcommand{\bea}[1]{\begin{eqnarray}\label{#1} }
\newcommand{\eea}{\end{eqnarray}}

\newcommand{\eqref}{\refb}

\usepackage{bm}
\usepackage[table]{xcolor}

\def\bj{{\bf j}}

\begin{document}

\baselineskip 24pt

\begin{center}

{\Large \bf Logarithmic Terms in the Soft Expansion in Four Dimensions}


\end{center}

\vskip .6cm
\medskip

\vspace*{4.0ex}

\baselineskip=18pt

\centerline{\large \rm Alok Laddha$^{a}$ and Ashoke Sen$^{b}$}

\vspace*{4.0ex}

\centerline{\large \it ~$^a$Chennai Mathematical Institute, Siruseri, Chennai, India}

\centerline{\large \it ~$^b$Harish-Chandra Research Institute, HBNI}
\centerline{\large \it  Chhatnag Road, Jhusi,
Allahabad 211019, India}


\vspace*{1.0ex}
\centerline{\small E-mail:  aladdha@cmi.ac.in, sen@hri.res.in}

\vspace*{5.0ex}

\centerline{\bf Abstract} \bigskip

It has been shown that in larger than four space-time dimensions, 
soft factors that relate the amplitudes with a soft photon or graviton to amplitudes
without the soft particle also determine the low frequency radiative part of the electromagnetic 
and gravitational fields during classical scattering. In four dimensions the S-matrix becomes
infrared divergent making the usual definition of the soft factor ambiguous beyond the
leading order. However the radiative
parts of the electromagnetic and gravitational fields  provide an unambiguous definition of soft factor in the 
classical limit up to the usual gauge
ambiguity.
We show that the soft factor
defined this way  develops terms involving logarithm of the energy of the soft particle at the subleading
order in the soft expansion.



\vfill \eject

\baselineskip 18pt

\tableofcontents

\section{Introduction and summary} \label{s1}

Soft theorems, that relate an amplitude with soft photons or gravitons to amplitudes without
any soft particle\cite{Gell-Mann,low,saito,burnett,bell,duca,
weinberg1,weinberg2,jackiw1,jackiw2,ademollo,shapiro},
have been investigated intensively in recent years\cite{1103.2981,1404.4091,1404.5551,1404.7749,1405.1015,
1405.1410,1405.2346,
1405.3413,1405.3533,1406.4172,1406.5155,1406.6574,1406.6987,
1406.7184,1407.5936,
1407.5982,1408.4179,1410.6406,1411.6661,1412.3699,
1502.05258,
1503.04816,1504.01364,1504.05558,1504.05559,
1505.05854,1507.00938,1507.08829,1507.08882,
1509.07840,1511.04921,
1512.00803,1601.03457,1604.00650,1604.03355,1604.02834,
1604.03893,1607.02700,1610.03481,1611.02172,1611.07534,1611.03137,
1702.02350,1702.03934,1703.00024,1705.06175,1706.00759,1707.06803}, 
partly due to their connection to 
asymptotic symmetries\cite{1312.2229,1401.7026,1408.2228,
1411.5745,1502.02318,1502.06120,1502.07644,1505.05346,
1506.05789,1509.01406,1605.09094,1605.09677,1608.00685,1612.08294,1701.00496,
1612.05886,1703.01351,1703.05448,1707.08016,
1709.03850,1711.04371,1712.01204,1712.09591,1803.03023}. 
Much of the discussion that relates soft theorem to asymptotic
symmetries has been in the context of four dimensional
theories, although there are some exceptions. 
However in four space-time dimensions the S-matrix suffers from infrared divergences
and is ill-defined. Therefore it is not obvious what soft theorem means in four space-time dimensions
beyond tree level. Indeed, 
in the case of gravity and abelian gauge theory, it has been shown that 
the leading soft factors are universal and are insensitive to infrared loop effects\cite{low,weinberg2},  
but
the subleading soft factors are infra-red divergent and can only be defined with 
appropriate regularization schemes\cite{1405.1015,1405.3413,1701.00496}.
For Yang-Mills theory, even the leading order 
soft factor is not universal once loop effects are taken into account and becomes regularization 
dependent\cite{1405.1015}.

In \cite{1801.07719} it was shown that in generic space-time 
dimensions, by taking classical limit of multiple soft
theorem, where we take the energies / charges of the finite energy external states to be
large, one can relate soft factors to the power spectrum of low frequency classical radiation  during a
scattering process. Since the latter can also be expressed in terms of the radiative components of the
electromagnetic and gravitational fields, this analysis yields a relation between the 
soft factors and the radiative components of 
low frequency electromagnetic and gravitational fields. 
If we take this relationship between classical radiation and soft theorem as the 
definition of the classical
soft factor\footnote{As soft factor beyond the leading order is a function of 
angular momenta of external states represented as differential operators, by classical soft factor 
we mean replacing these differential operators by classical angular momenta of external particles.}, 
it opens up the possibility of computing
the soft factor unambiguously by examining purely classical processes, even in four dimensions.
More precisely, in four space-time dimensions, 
the radiative components of the electromagnetic or gravitational field is given in terms of the soft factor
$S$ as
\be \label{e1.1}
-i\, {1\over 4\pi R} \, e^{i\omega R}\, S\, .
\ee

Thus for gravitational and electromagnetic fields, knowledge of the 
classical radiative field at the subleading order defines for us the 
corresponding soft factor at subleading order. 
We can then use this to explore the effect of infrared
divergences.
This is the task we undertake in this paper. We find that while the radiative part of classical fields is 
well defined in a classical scattering process, the problem appears when we try to carry out a Taylor
(more precisely Laurent) series expansion in the frequency $\omega$ of the soft radiation. 
The leading term of order $\omega^{-1}$
is well defined but at the subleading order there is a term proportional to $\ln\omega$ in four
dimensions. This dominates the order unity term that is usually the subleading soft factor
in higher dimensions. 

One can in fact find a trace of such logarithmic corrections in the standard soft theorem itself. Both
for electromagnetism and gravity, the subleading soft theorem has terms 
proportional to the angular momentum $\bj^{\mu\nu}$ of the incoming and outgoing finite energy
objects. For a classical particle with trajectory $r^\mu(\tau)$, where $\tau$ is the proper time,
the orbital part of $\bj^{\mu\nu}$ is given by $r^\mu p^\nu - r^\nu p^\mu$, where $p^\mu 
= m \, dr^\mu/d\tau$ and $m$ is the mass of the particle.
In dimensions
higher than four, $r^\mu$ grows as $V^\mu \tau + c^\mu$ for large $|\tau|$, where
$V^\mu$ and $c^\mu$ are constants. It is 
easy to see that $\bj^{\mu\nu}$ computed using this expression is $\tau$ independent and therefore has
a finite $\tau\to\pm\infty$ limit. However in four space-time dimensions, in the large $|\tau|$ limit, 
$r^\mu(\tau)$ will have an additional term proportional to $\ln |\tau|$ due to the long range
attractive force due to other particles involved in the scattering. It is easy to verify that $\bj^{\mu\nu}$
now acquires terms proportional to $\ln|\tau|$ which do not have finite limit as 
$\tau\to\pm\infty$. Therefore
the soft theorem itself shows that it breaks down in four space-time dimensions.

A naive guess would be that the logarithmic terms at the subleading order may be given simply by
replacing $\ln|\tau|$ by $\ln\omega^{-1}$ in the usual soft theorem. We set out to test this by examining
the explicit formula for radiative fields during classical scattering processes. 
We find that this is indeed true for
all cases for which we carry out the analysis.

We now give a summary of our results. The first scattering we analyze is that of a probe of a charge
$q$ and mass $m$ from a heavy scatterer of charge $Q$ and mass $M_0$ 
via electromagnetic interaction, and compute
the radiative part of the electromagnetic field of polarization $\ve$ and frequency 
$\omega$ along the direction $\hat n$. 
By comparing this with \refb{e1.1} 
we extract the soft factor in four dimensions. The result takes the form
\be\label{ethis}
\tilde S_{\rm em} =  - {q\over \omega} \left[
{\vec \ve.\vec \beta_+\over 1-\hat n.\vec\beta_+} - {\vec \ve.\vec \beta_-\over 1-\hat n.\vec\beta_-}
\right] 
- i \, q\, \ln \omega^{-1} \, 
\left[ C_+\,
{\vec \ve.\vec \beta_+\over 1-\hat n.\vec\beta_+} - 
C_-\, 
{\vec \ve.\vec \beta_-\over 1-\hat n.\vec\beta_-}
\right] + \hbox{finite}\, ,
\ee
where $\vec\beta_\pm$ denotes the velocities $d\vec r/dt$ of the probe as $t\to\pm\infty$, and
\be\label{ecpmemintro}
C_\pm = \pm  {q\, Q\over 4\pi \, m\, |\vec\beta_\pm|^3}\, (1-\vec\beta_\pm^2)^{3/2}\, .
\ee
\refb{ethis} agrees with what we would get by replacing the $\ln |\tau|$ factor 
in the soft theorem by $\ln\omega^{-1}$.

Next we analyze a similar scattering, but instead of computing emission of electromagnetic 
wave, we compute the emission of gravitational wave. However we ignore the effect of gravitational 
force on the scattering, treating gravity at the linearized level sourced by the energy density carried by
the probe and the electromagnetic field. By comparing this with \refb{e1.1} 
we extract the following form of the soft graviton
factor:
\ben \label{egrgivenintro}
S_{\rm gr} &=& -{m\over \omega} \, \ve^{ij}\, 
\left\{ {1\over 1 - \hat n.\vec \beta_+} \, {1\over \sqrt{1-\vec \beta_+^2}}\, \beta_{+ i} 
\beta_{+ j} - {1\over 1 - \hat n.\vec \beta_-} \, {1\over \sqrt{1-\vec \beta_-^2}}\, \beta_{- i} 
\beta_{- j}\right\}\nonumber \\
&& \hskip -.2in - i\, m \, \ln \omega^{-1} \, \ve^{ij}\, \left[ 
{1\over \sqrt{1-\vec \beta_+^2}}\, \beta_{+ i} 
\beta_{+ j} C_+ {1\over 1-\hat n.\vec\beta_+}  
- {1\over \sqrt{1-\vec \beta_-^2}}\, \beta_{- i} 
\beta_{- j} \, C_- {1\over 1-\hat n.\vec\beta_-} 
\right] + \hbox{finite} \, ,  \nonumber \\
\een
with $C_\pm$ given by \refb{ecpmemintro}. This also agrees with what one would get from the
soft theorem by replacing the $\ln |\tau|$ factor by $\ln \omega^{-1}$.

Our final example involves scattering of a neutral probe of mass $m$ from a massive scatterer of mass
$M_0$ via gravitational force in the limit of large impact parameter. For this analysis we take into
account the non-linear effects of gravity, {\it e.g.} the gravitational field produced by the 
probe and the scatterer acts as the source of gravity.  The soft graviton factor extracted
from this analysis takes the same form as \refb{egrgivenintro}
where now, in the $8\pi\, G=1$ unit,
\be 
C_\pm = \mp  {M_0 (1-3\vec\beta_\pm^2)\over 8\pi |\vec \beta_\pm|^3} .
\ee
This again agrees with what we would get by replacing $\ln |\tau|$ by
$\ln\omega^{-1}$ in the usual soft theorem.  

In the last example there is an additional subtlety that needs some discussion.
Since the long range
gravitational force acts on the soft graviton as well, the trajectory of the soft graviton far away from the 
scatterer takes the form
$t = R + (4\pi)^{-1} \, M_0 \, \ln R$. For this reason the radiative component of the
gravitational field will be proportional to $\exp[i\omega \{R + (4\pi)^{-1} \, M_0 \, \ln R- t\}]/R$ instead of the usual
factor
$\exp[i\omega (R-t)]/R$. 
Therefore \refb{e1.1} should contain
an infrared divergent phase factor of $\exp\left[i\omega \, {M_0 \over 4\pi} \ln R\right]$. 
To this end we would like to remind the
reader that the procedure for taking the classical limit, as described in \cite{1801.07719}, 
does not fix the overall
phase in \refb{e1.1}; this must be fixed by comparison with explicit results.
Comparison
with the results of explicit calculation shows
that the additional factor is
\be \label{ephase}
\exp\left[i\omega \, {M_0 \over 4\pi} \ln (\omega \, R)\right]\, .
\ee 
This phase factor, although present, is harmless since this does not affect the flux
of soft gravitons, although it can affect the shape of the gravitational wave-form. 
We expect this to be related to the infrared divergent corrections 
to the soft factor found in \cite{1405.1015}, and the classical counterpart of this calculation
given in \cite{0912.4254,1203.2962,1211.6095}. The term proportional to $\ln R$ represents the time delay
of a gravitational wave to reach its target at distance $R$ due to 
the long range gravitational force of the mass $M_0$.

Physically the corrections to the soft theorem associated with the $\ln |\tau|$ terms in $\bj^{\mu\nu}$ 
may be understood as the effect of the early and 
late time acceleration and deceleration of the finite energy particles due to the
long range force that they exert on each other. Due to this effect the particles continue
to radiate even at large time, producing soft radiation that is responsible for the $\ln\omega^{-1}$
contribution. 

It is natural to ask what these results mean for the quantum theory. As already pointed out, since the
S-matrix itself is divergent, in general the soft factor is ambiguous unless the divergences cancel from
both sides. The correct approach to studying soft theorem in four space-time dimensions would be to 
work with finite quantities like inclusive cross section\cite{bloch,kinoshita,lee} 
or Fadeev-Kulish 
formalism\cite{kulish,1308.6285,1705.04311}, 
and then see how
cross sections / amplitudes with and without soft external states are  related. Presumably by 
taking the classical limit of such modified multiple soft graviton theorem
as in \cite{1801.07719} we shall reproduce the results of this paper. 
Furthermore in that analysis the
terms proportional to $\ln \omega^{-1}$ would appear directly as $\ln\omega^{-1}$ and there will be no
need to make an ad hoc replacement of $\ln |t|$ by $\ln \omega^{-1}$. This has not been checked from 
first principles.

We end this section with a few remarks.
\begin{enumerate}
\item The soft factors given in \refb{ethis}, \refb{egrgivenintro}  
have finite $|\vec\beta_\pm|\to 1$ limit if we keep the energies
$E_\pm=m/\sqrt{1-\vec\beta_\pm^2}$ fixed in this limit.  Therefore the results are also valid for massless
probes.
\item The leading terms in the soft factors have the property that they vanish in the limit when the 
deflection goes to zero. This can be seen by setting $\vec\beta_+=\vec\beta_-$. This is also the
property of the usual subleading factors that arise in higher dimensions. In contrast, the logarithmic
terms in \refb{ethis}, \refb{egrgivenintro}  do not vanish in the limit $\vec\beta_+\to\vec\beta_-$
since $C_\pm$ have opposite signs. 
This is a reflection of the fact that the logarithmic terms come from the early and
late time acceleration due to
the long range force, and this persists even in the absence of any scattering.
\item For real polarizations, the terms proportional to
$\ln\omega^{-1}$ in \refb{ethis} and \refb{egrgivenintro} are purely imaginary. 
Therefore they do not contribute to
the power spectrum -- proportional to $|S|^2$ -- to subleading order. However for circular polarizations,
for which the $\ve$'s are complex, there may be non-vanishing 
contribution to the power spectrum at
the subleading order, since the tensors that are contracted with the polarizations at the leading and
subleading orders are different, and the subleading contribution cannot be factored out as a pure phase.
\item Our analysis also suggests a regime in parameter space where the usual soft expansion
may dominate the logarithmic terms. For definiteness let us focus on soft graviton emission.
If the scattering takes place via some interaction of range $b$ that is large
compared to the Schwarzschild radius $M_0/(4\pi)$  of the scatterer, then for impact parameter
of order $b$ and sufficiently 
large interaction strength -- {\it e.g.} hard elastic scattering -- we can
produce appreciable deflection.
This would give a leading contribution to $S_{\rm gr}$ of order $m/\omega$ and the usual subleading
soft factor of order $m\, b$ since the soft expansion parameter is of order $\omega\, b$.
On the other hand the logarithmic term is of order $m\, M_0\, \ln\omega^{-1}$. Therefore for $b>>M_0$
we can choose a range of $\omega$ in which the soft expansion parameter $\omega\, b$ is small,
but $b>> M_0\, \ln\omega^{-1}$. In this range the usual soft terms will dominate the logarithmic term.
Examples of such scattering can be found in sections 7.2.1 and 7.2.2 of \cite{1801.07719}.
\end{enumerate}

\sectiono{Logarithmic corrections from soft factors} \label{s2}

In this section we shall see that even the usual soft theorems -- valid in dimensions larger
than four -- develop
logarithmic factors when extrapolated to four space-time dimensions. We shall begin by reviewing the
results of \cite{1801.07719} that relates the soft factor to the radiative component of electromagnetic and
gravitational fields.  The general relation in $D$-dimensions
takes the following form for the gravitational field 
$\tilde h_{\alpha\beta}(\omega, \vec x)$, related to 
$h_{\alpha\beta}(t,\vec x)=(g_{\alpha\beta}-\eta_{\alpha\beta})/2$ by Fourier transform in 
the time variable:
\ben \label{ehabexp}
&& \tilde h_{\alpha\beta}(\omega,\vec x) = \tilde e_{\alpha\beta}(\omega,\vec x)
- {1\over D-2} \, \eta_{\alpha\beta}\, \tilde e_\gamma^{~\gamma}(\omega, \vec x)\, ,
\nonumber \\
&& \ve^{\alpha\beta}\, \tilde e_{\alpha\beta}(\omega, \vec x) = \NN'\,
S_{\rm gr}(\ve, k) \, ,
\nonumber \\
&& R \equiv |\vec x|, \quad \NN' \equiv  e^{i\omega R}\, \left({\omega\over 2\pi i R}\right)^{(D-2)/2} 
{1\over 2\omega}, \quad k \equiv -\omega(1, \hat n), \quad \hat n={\vec x\over |\vec x|}  \, .
\een
Here $\ve$ is any arbitrary rank two polarization 
tensor and $S_{\rm gr}$ is the soft factor for gravity whose expression will be given in 
\refb{essgr}. 
A similar formula exists for electromagnetism. The radiative component of the gauge field
$\wt A_\alpha(\omega, \vec x)$, related to the gauge field $A_\alpha(t,\vec x)$ 
by Fourier transformation in the time variable,
is given by
\be \label{e2.2}
\ve^\alpha \wt A_\alpha(\omega, \vec x) = \NN'\, 
S_{\rm em}(\ve, k) \, .
\ee

We 
shall now write down the explicit form of $S_{\rm em}(\ve, k)$ and $S_{\rm gr}(\ve, k)$ to subleading order.
For simplicity we shall consider the scattering of a pair of particles and work in the probe approximation 
where one of the objects (the probe) has mass much larger than the other (the scatterer). In this case we
have
\be \label{essem}
S_{\rm em}(\ve, k) = q \, \sum_{a=1}^2 (-1)^{a-1}\, {\ve . p_{(a)}\over k. p_{(a)}} 
+  i \sum_{a=1}^2 (-1)^{a-1} \, q \, {\ve_\nu k_\rho \over p_{(a)}. k}
\, \bj_{(a)}^{\rho\nu} + \hbox{non-universal}\, ,
\ee
\ben \label{essgr}
S_{\rm gr}(\ve, k)&=& \sum_{a=1}^2 \left[{\ve_{\mu\nu} p_{(a)}^\mu p_{(a)}^\nu\over p_{(a)}.k}
+ \ve_{00}{p_{(a)}.k\over (k^0)^2} + 2 \, \ve_{0\nu} {p_{(a)}^\nu\over k^0}\right]\nonumber \\
&& +
 i\, \sum_{a=1}^2 \left[\left\{
{\ve_{\mu\nu} p_{(a)}^\mu k_\rho \over p_{(a)}. k} 
+ {\ve_{\nu 0} k_{\rho}\over k^0}
\right\}\bj_{(a)}^{\rho\nu}
+\, {\bJ^{ji}\over M_0}\, \left\{ {\ve_{i0} \, k_j \, p_{(a)}. k \over (k^0)^2}
+{\ve_{i\nu} \, p_{(a)}^\nu \, k_j\over k^0} \right\}\right]\, .
\een
Here $p_{(1)}$ and $p_{(2)}$ are the  momenta of the probe before and
after the scattering and $q$ is the charge of the probe. 
The scatterer
is initially taken to be at rest, with mass $M_0$ and angular momentum
$\bJ$. 
The indices $i,j,\cdots$ run over spatial coordinates and the indices $\mu,\nu,\cdots$ run over
all space-time coordinates.
$\bj_{(1)}$ and $\bj_{(2)}$ are
the  angular momenta of the probe before and after the scattering,
measured with respect to the space-time point describing the location of the
center of momentum of the scatterer at some particular instant of time before the scattering.
All momenta and angular momenta are measured with the 
convention that they are counted with positive sign for ingoing and negative sign for
outgoing particles; for charges this is accounted for by the explicit $(-1)^a$ factors
in \refb{essem}.
The indices are raised and lowered by flat metric $\eta^{\mu\nu}$ and
$\eta_{\mu\nu}$  with mostly plus signature. The non-universal terms in the soft photon theorem appear
at the subleading order but they will not affect our analysis below.

For electromagnetic radiation the radiative part of the field satisfies the constraint equation 
$k^\alpha\wt A_\alpha=0$. This is reflected in the invariance of $S_{\rm em}(\ve,k)$ under 
$\ve^\mu\to \ve^\mu+k^\mu$.
Therefore $\wt A_i$ determines $\wt A_0$ and we can focus on the spatial components $\wt A_i$.
Consequently we can restrict $\ve^\alpha$ to have only spatial components. On the other hand the radiative part
of the gravitational field satisfies the constraint $k^\mu \tilde e_{\mu\nu}=0$, reflected in the invariance of
$S_{\rm gr}(\ve, k)$ under $\ve_{\mu\nu}\to \ve_{\mu\nu}+ \xi_\mu k_\nu + \xi_\nu k_\mu$. This allows us to determine
$\tilde e_{0\mu}$ in terms of the spatial components $\tilde e_{ij}$ and we can focus on the spatial components
$\tilde e_{ij}$. Consequently we can choose $\ve^{\mu\nu}$ to have only transverse components $\ve^{ij}$.
These may be summarized as:
\be \label{epolcon1}
\ve^0=0, \quad \ve^{0\rho}=0. 
\ee

Since both electrodynamics and gravity has gauge symmetries, we can determine the field configurations
only up to a choice of gauge. Consequently \refb{ehabexp} and \refb{e2.2} are valid only up to gauge
transformations:
\ben
&& \delta \tilde h_{\mu\nu}=k_\mu\xi_\nu + \xi_\mu k_\nu \quad \Rightarrow \quad
\delta \tilde e_{\mu\nu} = k_\mu\xi_\nu + \xi_\mu k_\nu - \xi.k \, \eta_{\mu\nu}\nonumber \\
&& \delta \wt A_\mu = \xi \, k_\mu\, ,
\een
for arbitrary parameters $\xi_\mu$ and $\xi$. Using these in \refb{ehabexp} and \refb{e2.2} we see that
the physical part of $S_{\rm gr}(\ve, k)$ and $S_{\rm em}(\ve,k)$ are contained in those choices of
polarization tensor / vector that satisfy
\be\label{epolcon2}
k_\mu \, \ve^{\mu\nu} - {1\over 2}\, k^\nu \ve^\rho_{~\rho}=0, \qquad k_\rho\, \ve^\rho=0\, .
\ee
Combining these with \refb{epolcon1} we get
\be \label{epolcon}
\ve^{0\rho}=0, \quad k_i\ve^{ij}=0, \quad \ve^i_{~i}=0, \qquad \ve^0=0, \quad k_i\ve^i=0\, .
\ee

Let $m$ be the mass of the probe particle and $\vec\beta_-$
and $\vec \beta_+$  be
its initial and final velocities. Then we have
\be
p_{(1)} = {m\over \sqrt{1-\vec \beta_-^2}} (1, \vec \beta_-), \quad p_{(2)} = -{m\over \sqrt{1-\vec \beta_+^2} }
(1, \vec \beta_+)\, .
\ee
The minus sign in the expression for $p_{(2)}$ is a reflection of the fact that it is an outgoing momentum.
Of special interest will be the initial and final trajectories $r_{(1)}(t)$ and $r_{(2)}(t)$. 
In dimensions $D>4$ these can be taken to be
of the form
\be\label{etrtr1}
r_{(1)}^0=t, \quad r_{(2)}^0=t, \quad \vec r_{(1)} = \vec\beta_- t + \vec c_-, \quad 
\vec r_{(2)} = \vec\beta_+ t + \vec c_+\, ,
\ee
for constant vectors $\vec c_\pm$.
Therefore we have 
\ben
&& \hskip -.3in \bj_{(1)}^{ij} = r_{(1)}^i p_{(1)}^j - r_{(1)}^j p_{(1)}^i = {m\over \sqrt{1-\vec \beta_-^2}} (c_-^i \beta_-^j
- c_-^j \beta_-^i), \quad \bj_{(1)}^{0i} = r_{(1)}^0 p_{(1)}^i - r_{(1)}^i p_{(1)}^0=
-  {m\over \sqrt{1-\vec \beta_-^2}} c_-^i\, , \nonumber \\ && \hskip -.3in
\bj_{(2)}^{ij} = r_{(2)}^i p_{(2)}^j - r_{(2)}^j p_{(2)}^i = -{m\over \sqrt{1-\vec \beta_+^2}} (c_+^i \beta_+^j
- c_+^j \beta_+^i),\quad \bj_{(2)}^{0i} = r_{(2)}^0 p_{(2)}^i - r_{(2)}^i p_{(2)}^0=
 {m\over \sqrt{1-\vec \beta_+^2}} c_+^i\, .\nonumber \\
\een
In particular these approach finite limit as $t\to\pm\infty$. However in $D=4$ there is  a long range force on
the incoming and outgoing probe that falls off according to inverse square law. It is easy to verify that in this
case the particle trajectories \refb{etrtr1} are modified to 
\be\label{etrtr2}
r_{(1)}^0=t, \quad r_{(2)}^0=t, \quad \vec r_{(1)} = \vec\beta_- t + \vec c_- - C_- \, \vec\beta_-
\ln |t|, \quad 
\vec r_{(2)} = \vec\beta_+ t + \vec c_+- C_+\, \vec\beta_+ \, \ln |t|\, ,
\ee
for appropriate constants $C_\pm$. This modifies the expressions for $\bj_{(i)}^{\mu\nu}$ to
\ben
&& \hskip -.3in \bj_{(1)}^{ij} = {m\over \sqrt{1-\vec \beta_-^2}} (c_-^i \beta_-^j
- c_-^j \beta_-^i), \quad \bj_{(1)}^{0i} =
-  {m\over \sqrt{1-\vec \beta_-^2}} \left\{ c_-^i - C_-\, \beta_-^i\, \ln |t|\right\}\, , \nonumber \\ && \hskip -.3in
\bj_{(2)}^{ij} = -{m\over \sqrt{1-\vec \beta_+^2}} (c_+^i \beta_+^j
- c_+^j \beta_+^i),\quad \bj_{(2)}^{0i} =
{m\over \sqrt{1-\vec \beta_+^2}} \left\{c_+^i - C_+\, \beta_+^i \, \ln |t|\right\}\, .\nonumber \\
\een
Note in particular that $\bj_{(a)}^{0j}$ diverges as $|t|\to\infty$, making the expressions ill-defined. 
Ignoring this for the time being, for the particle kinematics described above
we can express the soft factors given in \refb{essem} and \refb{essgr}
as 
\be \label{e20pre}
S_{\rm em} =  - {q\over \omega} \left[
{\vec \ve.\vec \beta_+\over 1-\hat n.\vec\beta_+} - {\vec \ve.\vec \beta_-\over 1-\hat n.\vec\beta_-}
\right] - i \, q\, \ln |t| \, 
\left[C_+{\vec \ve.\vec \beta_+\over 1-\hat n.\vec\beta_+} - C_-
{\vec \ve.\vec \beta_-\over 1-\hat n.\vec\beta_-}
\right] + \hbox{finite}\, ,
\ee
and 
\ben \label{egrgiven}
S_{\rm gr} &=& -{m\over \omega} \, \ve^{ij}\, 
\left\{ {1\over 1 - \hat n.\vec \beta_+} \, {1\over \sqrt{1-\vec \beta_+^2}}\, \beta_{+ i} 
\beta_{+ j} - {1\over 1 - \hat n.\vec \beta_-} \, {1\over \sqrt{1-\vec \beta_-^2}}\, \beta_{- i} 
\beta_{- j}\right\}\nonumber \\
&& \hskip -.2in - i\, m \, \ln |t| \, \ve^{ij}\, \left[ 
{1\over \sqrt{1-\vec \beta_+^2}}\, \beta_{+ i} 
\beta_{+ j} C_+ {1\over 1-\hat n.\vec\beta_+}  
- {1\over \sqrt{1-\vec \beta_-^2}}\, \beta_{- i} 
\beta_{- j} \, C_- {1\over 1-\hat n.\vec\beta_-} 
\right] + \hbox{finite} \, ,  \nonumber \\
\een
where `finite' refers to terms which remain finite as $\omega\to 0$, $|t|\to\infty$.

A natural
guess is that the presence of $\ln |t|$ term  
implies the breakdown of the expansion of the soft factor in power series in 
$\omega$. Naively one might expect 
that the correct expression is given by replacing the $\ln |t|$ factors
by $\ln\omega^{-1}$. In the following we shall verify this by explicit computation in
several examples.

\sectiono{Some relevant integrals} \label{s3}

In our analysis we shall often encounter integrals of the form
\be \label{ei1}
I=\int dt \, e^{i\omega g(t)} F(t) + \hbox{boundary terms}\, ,
\ee
where $g(t)$ and $F(t)$ are functions of $t$ and the integration over $t$ runs from $-\infty$ to $+\infty$.
As will be discussed shortly, the `boundary terms' need to be adjusted to make the integral well-defined.
In all the examples considered, $g(t)$ will grow  as $a_\pm \, t$ as $t\to\pm\infty$
for some constants $a_\pm$, with 
possible corrections of order $\ln|t|$. $F(t)$ will typically either approach a constant or fall off as some 
negative power of $t$, again with possible subleading corrections involving $\ln |t|$. If 
$F(t)\sim |t|^{-\alpha}$ for $\alpha>0$, then the integral is well defined, and can be evaluated by
taking the limits to be from $-T$ to $T$ and taking the $T\to\infty$ limit. If on the other hand 
$F(t)\sim |t|^{-\alpha}$ with $-1< \alpha\le 0$, then we have to define the integral by first performing an
integration by parts:
\be \label{ei2}
I = \int dt\, \left\{{d\over dt} e^{i\omega g(t)}\right\} {1\over i\omega \, g'(t)} \, F(t)
+\hbox{boundary terms}
= -{1\over i\omega} \int dt\, e^{i\omega g(t)} \, {d\over dt} \left\{ {F(t)\over g'(t)} \right\}\, ,
\ee 
where the boundary terms have been chosen to cancel the boundary terms 
arising from integration by parts.
Since $g'(t)\to a_\pm$ as $t\to\pm\infty$ and $F(t)\sim |t|^{-\alpha}$ as $t\to\pm\infty$, the
integrand in \refb{ei2} falls off as $t^{-\alpha-1}$ and therefore for $\alpha>-1$ this can be
defined by putting limits $\pm T$ on the $t$ integral and then taking the limit $T\to\infty$. 
This makes the integral well-defined without boundary terms.
If $\alpha\le -1$ then
we need to carry out more integration by parts but we shall not encounter such a situation. 

Once we have defined the integral 
so that it can be evaluated as limits of an integral with finite range, we are free to go back to the
original form by integration by parts, but now we have to keep track of the boundary terms. This reduces
the integral to
\be \label{ei3}
I=\lim_{T\to\infty}\left\{
\int_{-T}^T dt \, e^{i\omega g(t)} F(t) -{1\over i\omega} \left[e^{i\omega g(t)} {F(t)\over g'(t)}
\right]_{-T}^T\right\}\, .
\ee
We can use either the right hand side of
\refb{ei2} or \refb{ei3} as the proper definition of \refb{ei1} after taking the
$T\to\infty$ limit.
The first term
in \refb{ei3} has less powers of $\omega$ in the denominator compared to the right hand side of
\refb{ei2}, but the boundary terms carry powers of $\omega$ in the denominator and provide the
missing terms. Therefore \refb{ei3} is convenient for carrying out a small $\omega$ expansion.
This can be done by first carrying out the small $\omega$ expansion of \refb{ei3} and then
taking the limit $T\to\infty$. This is
the strategy that was used in \cite{1801.07719} for checking soft theorems in dimensions $D>4$. 

This procedure is useful if the expansion does not have terms of order $\ln\omega^{-1}$, but runs
into difficulty if there are $\ln\omega^{-1}$ terms in the expansion. To see how this happens, note
that the right hand side of \refb{ei2} is 
well defined as integral over a finite range $(-T, T)$ and also in the limit $T\to\infty$. Therefore 
\refb{ei3} is also
well-defined for $T\to\pm\infty$. However suppose that 
we take the form given in \refb{ei3} and carry out the expansion in
$\omega$ before taking the $T\to\infty$ limit, and then in each term in the expansion take the $T\to
\infty$ limit. Since this will always produce power series in $\omega$, the only way we can see the
presence of the $\ln\omega^{-1}$ term is that the expansion coefficients will now fail to have finite
limit as $T\to\infty$ even though the original expressions \refb{ei2} and
\refb{ei3} have well defined $T\to\infty$ limit.

We shall analyze the logarithmic terms in the soft factor by always working with the convergent 
form of the integral as in the right hand side of
\refb{ei2} without explicit boundary terms and then analyzing the behavior
of the integral in the $\omega\to 0$ limit without first naively expanding the integrand in powers of
$\omega$. Below we write down the expressions of five different types of integrals that we shall
need for our analysis and the values of the integrals for small $\omega$. The derivation
can be found in appendix \ref{sa}.

Let $f,g,h$ and $r$ be functions of $t$ with the following asymptotic behavior:
\ben \label{e1}
&&  f(t) \to f_\pm+{k_\pm\over t}, \quad 
g(t) \to a_\pm t + b_\pm \ln |t|, \nonumber \\ &&
h(t) \to p_\pm t + q_\pm \ln |t|, \quad r(t)\to c_\pm \, t+ d_\pm \ln |t|, 
\quad \hbox{as} \quad t\to\pm\infty\, .
\een
In appendix \ref{sa} we prove the following results for arbitrary constant $R$:
\ben\label{e2}
I_1&\equiv& {1\over \omega} \int_{-\infty}^\infty dt \, e^{-i\,\omega\, g(t) } f'(t) 
= \omega^{-1} (f_+-f_-) + i\, (a_+k_+ - a_- k_-) \ln \omega^{-1} + \hbox{finite}\, ,\nonumber \\ \cr
I_2&\equiv&  \int_{-\infty}^\infty dt \,e^{-i\,\omega\, g(t) }  {d\over dt} \left[f(t) \left\{
\ln {h(t)\over R}+ \int_{h(t)}^\infty  e^{i\, \omega\, u} \, {du\over u}\right\}\right] = 
- (f_+-f_-) \ln (R\, \omega) +
\hbox{finite}\, ,
\nonumber \\ \cr
I_3&\equiv& \int_{-\infty}^\infty dt \, {1\over r(t)} \, f(t)\, \left[e^{-i\, \omega\, g(t)}-
e^{-i\, \omega\, h(t)}\right] = \hbox{finite}\, , \nonumber \\ \cr
I_4 &\equiv& {1\over \omega} 
\int_{-\infty}^\infty dt \, {1\over r(t)^2} \, f(t)\, \left[e^{-i\, \omega\, g(t)}-
e^{-i\, \omega\, h(t)}\right] \nonumber \\
&=&
- i \, \{f_+\, c_+^{-2}(a_+ - p_+) - f_-\, c_-^{-2}(a_- - p_-)\} \, \ln \, \omega^{-1} + \hbox{finite}\, ,
\nonumber \\ \cr
I_5 &\equiv& \int_{-\infty}^\infty dt \, {1\over r(t)} \, f(t)\, e^{-i\, \omega\, g(t)}
=(f_+ c_+^{-1} - f_- c_-^{-1}) \, \ln \omega^{-1} +\hbox{finite}\, .
\een

\sectiono{Electromagnetic radiation}

In this section we shall analyze the electromagnetic radiation due to the scattering of a charged probe from a
charged scatterer. This is given by\cite{1801.07719}
\be\label{ereturn}
\wt A_\alpha(\omega, \vec x) 
= i\, \NN'\,  \int d\sigma 
e^{i\omega \{r^0(\sigma) - \hat n.\vec r(\sigma)\}} \, q \, V_\alpha(\sigma) + \hbox{boundary terms}\, ,
\ee
where the `boundary terms' are determined using the principle described in section \ref{s3},
$\hat n=\vec x/|\vec x|$, $\sigma$ denotes the proper time along the particle trajectory
$r(\sigma)$ and $V^\alpha$
is the four velocity
\be 
V^\alpha(\sigma) = {d r^\alpha\over d\sigma} \, .
\ee
It can be shown that\cite{1801.07719}
in four space-time dimensions \refb{ereturn}
reduces to the standard formula for electromagnetic radiation from an accelerated particle described 
{\it e.g.} in \cite{jackson}. By a change of variables from $\sigma$ to $t=r^0$ and an integration by parts,
we can bring this expression into the form
\be\label{e4.3}
\wt A_i(\omega, \vec x) 
= -q\, \omega^{-1} \, \NN'\,  \int dt\, 
e^{i\omega \{t - \hat n.\vec r(t)\}} \, {d\over dt} \left\{ {1\over 1-\hat n.\vec v(t)} \, v_i(t)\right\}\, ,
\ee
where we have focussed on the spatial components of $\wt A$. Since it follows from \refb{etrtr2} that
$\vec v(t)=d\vec r/dt$ approaches a constant plus terms of order $1/|t|$ for large $t$, the integrand in
\refb{e4.3} falls off as $1/t^2$ and therefore we do not need to add boundary terms in this representation.

Comparing \refb{e4.3} with \refb{e2.2} we can identify the prediction for the soft factor from classical
analysis:
\be \label{e16}
\tilde S_{\rm em}(\ve, k) = - {q\over \omega} \int_{-\infty}^\infty \, dt \, e^{i\omega (t-\hat n.\vec r(t))}\, 
{d\over dt} \left[ {\vec \ve .\vec v\over 1-\hat n.\vec v}\right]\, ,
\ee
assuming that $\ve$ has only spatial components.
Now as $t\to\pm \infty$, we have
\be\label{e20a}
\vec r(t) \to \vec \beta_\pm t - C_\pm \vec\beta_\pm \ln |t| + \hbox{finite},
\quad \vec v(t) \equiv {d\vec r(t)\over dt} \to \vec\beta_\pm (1 - C_\pm t^{-1}).
\ee
Therefore the integral in \refb{e16} has the form of $I_1$ given in \refb{e2} with
\ben \label{e18}
g(t) &\equiv& 
\hat n.\vec r(t) - t \simeq (\hat n.\vec\beta_\pm - 1) \, t - C_\pm \, \hat n.\vec\beta_\pm \, \ln |t| + \hbox{finite},
\nonumber \\
f(t) &\equiv& -q \, \left[ {\vec \ve .\vec v(t)\over 1-\hat n.\vec v(t)}\right]
= -q {\vec \ve.\vec \beta_\pm\over 1-\hat n.\vec\beta_\pm} \left[1 - {C_{\pm}\over 1-\hat n.\vec\beta_\pm} t^{-1}
+\cdots\right]\, , \quad \hbox{as $t\to\pm\infty$}\, .
\een
Comparing  \refb{e18} with \refb{e1} we get
\be 
a_\pm = (\hat n.\beta_\pm - 1), \quad f_\pm = 
-q {\vec \ve.\vec \beta_\pm\over 1-\hat n.\vec\beta_\pm} , \quad k_\pm = 
q \, C_\pm \, {\vec \ve.\vec \beta_\pm\over (1-\hat n.\vec\beta_\pm)^2} \, .
\ee
Therefore we get from \refb{e2}
\be \label{e20}
\tilde S_{\rm em} =  - {q\over \omega} \left[
{\vec \ve.\vec \beta_+\over 1-\hat n.\vec\beta_+} - {\vec \ve.\vec \beta_-\over 1-\hat n.\vec\beta_-}
\right] - i \, q\, \ln \omega^{-1} \, 
\left[C_+{\vec \ve.\vec \beta_+\over 1-\hat n.\vec\beta_+} - C_-
{\vec \ve.\vec \beta_-\over 1-\hat n.\vec\beta_-}
\right] + \hbox{finite}\, .
\ee
This agrees with $S_{\rm em}$ given in \refb{e20pre} if we replace $\ln |t|$ by $\ln\omega^{-1}$.

For completeness let us compute $C_\pm$. Asymptotically we can regard the velocity 
carried by the probe to be
in the radial direction. If $Q$ denotes the charge carried by the scatterer, then energy conservation gives
\be
{m\over \sqrt{1-\vec v(t)^2}} +  {q\, Q\over4\pi |\vec r(t)|}= \hbox{constant} \, .
\ee
Substituting \refb{e20a} into this equation and setting the coefficient of the $1/|t|$ term for large $|t|$
to zero, we get
\be\label{ecpmem}
C_\pm = \pm  {q\, Q\over 4\pi \, m\, |\vec\beta_\pm|^3}\, (1-\vec\beta_\pm^2)^{3/2}\, .
\ee

\sectiono{Gravitational radiation}

In this section we shall analyze the logarithmic correction to the soft factor for gravitational radiation.
We shall analyze two examples. In the first the scattering takes place via electromagnetic interaction
and the energy momentum tensor during the scattering is used as a source for gravitational radiation.
Assuming that the electromagnetic interaction is much stronger than the gravitational interaction during
the scattering, we ignore the effect of gravity on the motion of the probe. Therefore
for this problem, the non-linear effects of gravity are suppressed. The second example involves the
scattering of a neutral probe off a massive object via gravitational interaction. For this problem the
non-linear effects of gravity become important since the gravitational field itself acts as
a source of gravitational radiation.

\subsection{Gravitational radiation from scattering via electromagnetic interaction}

The set up here is as follows. The probe has mass $m$ and charge $q$ and the scatterer has mass 
$M_0>>m$ 
and
charge $Q>>q$. We assume that the distance of closest approach between 
the probe and the scatterer is large
compared to the Schwarzschild radius of  the scatterer so that the effect of gravity on the scattering can be
ignored, but that $Q$ and $q$ are sufficiently large so that there is appreciable scattering due to the 
electromagnetic force. In this case the energy momentum tensor, that acts as the source of gravitational 
radiation, receives contribution from two sources -- the probe and the electromagnetic field. As long as we 
focus on the spacial components of $\tilde e_{ij}$ whose source is the spatial component of the energy momentum
tensor, we can ignore the contribution due to the scatterer due to the smallness of its velocity during the scattering.
Consequently 
the result for $\tilde e_{ij}$ is given by the sum of two terms: $\tilde e^{(1)}_{ij}$ due to the probe and
$\tilde e^{(2)}_{ij}$ due to the electromagnetic field. We shall now analyze each component separately.

The radiative part of the gravitational field due to the probe is given by\cite{1801.07719}
\be \label{ex1}
\tilde e^{(1)}_{\ia\ja} =  i \, \NN'\, \int d\sigma \, e^{ik.r(\sigma)} \, P_\ia(\sigma)\, V_\ja(\sigma)
+ \hbox{boundary terms} \, ,
\ee
where the integral runs over the proper time $\sigma$ along the world-line $r(\sigma)$ of the probe, 
$V^\alpha = dr^\alpha/d\sigma$ is the $D$-velocity of the probe and $P^\alpha=m V^\alpha$ is the
momentum of the probe. We now change the integration variable from $\sigma$ to $t=r^0$ to
express \refb{ex1} as
\be \label{ex11}
\tilde e^{(1)}_{\ia\ja} =  i \, \NN' \, \int dt \, e^{ik.r} \left({dt\over d\sigma}\right)^{-1} \, P_\ia\, V_\ja
+ \hbox{boundary terms} 
\, .
\ee
Using 
\be
(k^0, \vec k) =-\omega (1, \vec n)\, ,
\ee
\be 
V^0 = {dr^0\over d\sigma} = {dt\over d\sigma} = {1\over \sqrt{1-\vec v^2}}, 
\quad \vec v(t) \equiv {d\vec r\over dt}\, ,
\ee
\be 
\vec V = {d\vec r\over d\sigma} ={dt\over d\sigma} \, {d\vec r\over dt}  
= {1\over \sqrt{1-\vec v^2}} \vec v(t)\, ,
\ee
and 
\be 
\vec P = m \, \vec V = {m\over \sqrt{1-\vec v^2}} \, \vec v\, ,
\ee
we can express \refb{ex11} as
\ben \label{egiven}
\tilde e^{(1)}_{ij} &=&
i \, \NN' \, m\, 
\int dt \, e^{i\omega (t - \vec n.\vec r(t))} 
{1\over \sqrt{1-v^2}} \, v_i v_j 
+ \hbox{boundary terms} \nonumber \\
&=& i \, \NN' \, m\, \int dt  \, {1\over i\omega (1 -\vec n.\vec v)} {d\over dt}
\left(  e^{i\omega (t - \vec n.\vec r(t))} \right) {1\over \sqrt{1-v^2}} \, v_i v_j 
+ \hbox{boundary terms} \nonumber \\
&=& - \NN'\ \, m\, \omega^{-1} \int dt \, e^{i\omega (t - \vec n.\vec r(t))} 
\, {d\over dt} \left\{ {1\over (1 -\vec n.\vec v)} {1\over \sqrt{1-v^2}} \, v_i v_j \right\}\, .
\een
Note that since $\vec v$ approaches a constant plus terms of order $1/|t|$ for large $t$, the
integral in the last line is convergent and we do not need to add any boundary terms. In this case
$\vec r(t)$ has the form given in \refb{e20a}
as $t\to\pm\infty$, 
with $C_\pm$ 
given by \refb{ecpmem}, since in our approximation the long range force between the probe and the scatterer is purely
electromagnetic.
Therefore in these limits,
\ben
&&  {1\over 1 - \hat n.\vec v(t)}  \, {1\over \sqrt{1-\vec v^2}}\, v_i v_j
\nonumber \\ &\simeq& {1\over 1 - \hat n.\vec \beta_\pm} \, {1\over \sqrt{1-\vec \beta_\pm^2}}\, \beta_{\pm i} 
\beta_{\pm j}
\left[ 1  - {1\over t} \left\{C_\pm {\hat n.\vec\beta_\pm\over 1-\hat n.\vec\beta_\pm} +C_\pm {\vec\beta_\pm^2\over 1 -\vec\beta_\pm^2} 
+2 \, C_\pm 
\right\}\right]\, , \nonumber \\ &&
t - \hat n. \vec r(t))\simeq t(1 -\hat n.\vec \beta_\pm) + C_\pm \hat n.\vec \beta_\pm \, \ln |t|\, .
\een
Therefore we can use the formula for $I_1$ to express $\tilde e^{(1)}_{ij}$ given in \refb{egiven} as
\ben \label{ex1.9}
\tilde e^{(1)}_{ij} &=& -\,\omega^{-1} \, m\, \NN' \, 
\left\{ {1\over 1 - \hat n.\vec \beta_+} \, {1\over \sqrt{1-\vec \beta_+^2}}\, \beta_{+ i} 
\beta_{+ j} - {1\over 1 - \hat n.\vec \beta_-} \, {1\over \sqrt{1-\vec \beta_-^2}}\, \beta_{- i} 
\beta_{- j}\right\}\nonumber \\
&& - i\, m\, \NN'\,   \ln \omega^{-1} \left[ 
{1\over \sqrt{1-\vec \beta_+^2}}\, \beta_{+ i} 
\beta_{+ j} \left\{C_+ {1\over 1-\hat n.\vec\beta_+}  +C_+ {1\over 1 -\vec\beta_+^2} 
\right\} \right.\nonumber \\ &&
\left. - {1\over \sqrt{1-\vec \beta_-^2}}\, \beta_{- i} 
\beta_{- j} \left\{C_- {1\over 1-\hat n.\vec\beta_-} 
+C_- {1\over 1 -\vec\beta_-^2} 
\right\}
\right]\, .
\een

We now turn to the contribution $\tilde e^{(2)}$ produced by the electromagnetic field. The dominant
part of the stress tensor comes from the term proportional to $Q^2$, but since the electric field produced
by the scatterer is stationary, it does not generate any radiative component. Therefore we focus on the
next term proportional to $q\, Q$. Denoting by $F^P_{\mu\nu}(x)$ and $F^S_{\mu\nu}(x)$ 
the field strengths
produced by the probe and scatterer respectively, and by $\wt F^P_{\mu\nu}(\omega,\vec \ell)$ 
and $\wt F^S_{\mu\nu}(\omega,\vec \ell)$ their Fourier transform in the space and time variables, 
we have\cite{1801.07719}
\be \label{e775a}
\wt F^S_{i0}(\ell) = -i\, \ell_i\, Q \, {1\over \vec \ell^2} \, 2\pi \, \delta(\ell^0)\, ,
\ee 
and
\be \label{e776}
\wt F^P_{i0}(-\ell) =-q\, 
{1\over (\ell^0-i\eps)^2 - \vec \ell^2} \, \int d\sigma \, e^{i \ell. r(\sigma)}
\left\{ -i \, \ell_i \, {dr_0\over d\sigma} + i \, \ell_0 \, {dr_i\over d\sigma} \right\}\, .
\ee
Then in $D$ dimensions $\tilde e^{(2)}_{ij}$  is given by\cite{1801.07719}
\ben \label{ee2def}
\tilde e^{(2)}_{ij} &=&  i \, \NN'\, \int d^D x'  \, e^{ik.x'}
\,  \left[- F^P_{i 0}(x') \,  F^S_{j0}(x') 
-  F^P_{j 0}(x') \,  F^S_{i0}(x') 
+
 \delta_{ij} \, F^S_{k0}(x') 
\,  F^{P}_{k0}(x')\right] \nonumber \\
&&\hskip -.6in =  i \, \NN'\, \int {d^D\ell\over (2\pi)^D} 
\,  \Bigg[- \wt F^P_{i 0}(-\ell-k) \,  \wt F^S_{j0}(\ell) 
-  \wt F^P_{j 0}(-\ell-k) \,  \wt F^S_{i0}(\ell) 
+
 \delta_{ij} \, \wt F^P_{k0}(-\ell-k) 
\,  \wt F^{S}_{k0}(\ell)\Bigg]\, .
\nonumber \\
\een
Using \refb{e775a}, \refb{e776} this may be rewritten as
\ben \label{exx77}
\tilde e^{(2)}_{ij} &=&  i \, \NN' \int d\sigma \int {d^{D-1}\ell\over (2\pi)^{D-1}} \, e^{i \vec \ell.
\vec r(\sigma) + i k . r(\sigma)}
\,  q\,  Q\, {1\over (\vec\ell^2)(\vec\ell^2 + 2\vec\ell.\vec k)} 
\nonumber \\ && 
\left[\left\{2\, \ell_i \ell_j +\ell_i k_j + \ell_j k_i 
-(\vec\ell^2 +\vec \ell . \vec k) \, \delta_{ij}\right\} \, 
{d r_0\over d\sigma} + \left\{ - k_0 \ell_j {d r_i\over d\sigma} - k_0 \, \ell_i \, {d r_j\over d\sigma} 
+ k_0 \, \ell_m \, {d r_m\over d\sigma} \, \delta_{ij}
\right\}
\right]\nonumber \\
&& \hskip -.3in 
= i \, \NN'\, \int d\sigma \int {d^{D-1}\ell\over (2\pi)^{D-1}} \, e^{i \vec \ell.
\vec r(\sigma) + i k . r(\sigma)}
\,  q\,  Q\, {1\over (\vec\ell^2)^2 }
\{2\, \ell_i \ell_j 
-\vec\ell^2  \, \delta_{ij}\} \, 
{d r_0\over d\sigma}  + \tilde f_{ij}\, , \een
\ben \tilde f_{ij} 
&\equiv&   i \, \NN'\, \int d\sigma \int {d^{D-1}\ell\over (2\pi)^{D-1}} \, e^{i\vec \ell.
\vec r(\sigma)+ i k . r(\sigma)}
\,  q\,  Q\, {1\over (\vec\ell^2)^2 (\vec\ell^2 + 2\vec\ell.\vec k)} \nonumber \\ &&
\hskip 1in \left\{-4\, \ell_i \ell_j \, \vec\ell . \vec k + \vec\ell^2 (\ell_i k_j + \ell_j k_i) 
+ \vec \ell. \vec k\, \vec\ell^2  \, \delta_{ij}\right\} \, 
{d r_0\over d\sigma} \nonumber \\
&&  + \, i \, \NN'\, \int d\sigma \int {d^{D-1}\ell\over (2\pi)^{D-1}} \, e^{i\vec \ell.
\vec r(\sigma)+ i k . r(\sigma)}
\,  q\,  Q\, {1\over (\vec\ell^2) (\vec \ell^2 + 2\vec \ell.\vec k)} \nonumber \\ && \hskip 1in
\, \left\{ - k_0 \ell_j {d r_i\over d\sigma} - k_0 \, \ell_i \, {d r_j\over d\sigma} 
+ k_0 \, \ell_m \, {d r_m\over d\sigma} \, \delta_{ij}
\right\}
\, . 
\een

In the expression for $\tilde f_{ij}$ the integration over $\ell$ is free from infrared divergence for
$D\ge 4$ even 
after we factor out a power of $\omega$ and then take the
$k\to 0$ limit. Furthermore $dr_0/d\sigma$ and $dr_i/d\sigma$ approach finite values as $\sigma\to
\pm\infty$. Taking into account the explicit factor of $k$ in all the terms in $\tilde f_{ij}$, we have
\be
\tilde f_{ij} = \omega \int dt e^{ik.r(t)} \, f(t) + \hbox{boundary terms}
\ee
where $f(t)$ approaches a finite value as $t\to\pm\infty$. Rewriting this as
\be
- \int dt \,  e^{ik.r(t)} \, {d\over dt} \left\{ {1\over i(1-\hat n.\vec v)} f(t)\right\}\, ,
\ee
we see that this has the form $\omega I_1$. Therefore it does not have any divergent contribution
in the $\omega\to 0$ limit and we can focus on the contribution to $\tilde e^{(2)}_{ij}$ from the
first term on the right hand side of \refb{exx77}. 

Using 
\be
{1\over (\vec\ell^2)^2} \{2\, \ell_i \ell_j -\vec\ell^2 \, \delta_{ij}\}
= -{1\over 2} \left[{\p \over \p \ell_i} \left({\ell_j\over \vec\ell^2}\right) 
+ {\p \over \p \ell_j} \left({\ell_i\over \vec\ell^2}\right)
\right], 
\ee
and integration by parts, we can express \refb{exx77} as
\ben\label{elast2}
\tilde e^2_{ij} &\simeq& 
{i\over 2} \, \NN'\, \int d\sigma \int {d^{D-1}\ell\over (2\pi)^{D-1}} \, e^{i \vec \ell.
\vec r(\sigma) +ik.r(\sigma)}
\,  q\,  Q\, {1\over \vec\ell^2 }
\{i\ell_i r_j + i\ell_j r_i\} \, 
{d r_0\over d\sigma} \nonumber \\
&=& - {i\over 2} \, \NN'\,  \int d\sigma \int {d^{D}\ell\over (2\pi)^{D}} \, e^{i  \ell.
r(\sigma) +ik.r(\sigma)}
\,  
\{q\, \wt F^S_{i0}(\ell) \, r_j + q\, \wt F^S_{j0}(\ell) \, r_i\} \, 
{d r_0\over d\sigma} \nonumber \\
&=& - {i\over 2} \, \NN'\, \int d\sigma\,  e^{ik.r(\sigma)}\, \{q\, F^S_{i0}(r(\sigma)) \, r_j(\sigma) + q\,
F^S_{j0}(r(\sigma)) \, 
r_i(\sigma)\} {dr_0\over d\sigma} \, ,
\een
where in the second step we have used \refb{e775a}.
Using equations of motion
\be \label{efs1}
{dP_\alpha\over d\sigma} = q \,  F^S_{\alpha\rho}(r(\sigma)) \, {d r^\rho\over d\sigma}\, ,
\ee
and the identification $r_0=-r^0=-t$,
we can express \refb{elast2} as
\ben 
\tilde e^{(2)}_{ij} &=& {i\over 2} \, \NN'\, \int d\sigma \, e^{ik.r(\sigma)} \, \left\{ {dP_i\over d\sigma} \, r_j(\sigma) 
+ {dP_j\over d\sigma} \, r_i(\sigma)  \right\} \nonumber \\
&=& i\, {m\over 2} \, \NN'\,  \int dt\,  e^{ik.r}\, \left[{d\over dt}\left\{
{v_i\over \sqrt{1-\vec v^2}}
\right\} \, r_j + 
{d\over dt}\left\{
{v_j\over \sqrt{1-\vec v^2}}
\right\} \, r_i\right]\, .
\een
Now specializing to the case $D=4$ and 
using \refb{e20a} we see that the term inside the square bracket 
behaves in the limit $t\to\pm\infty$, as
\be 
2\, t^{-1}\, 
C_\pm \, \beta_{\pm i} \, \beta_{\pm j} \, {1\over (1-\vec\beta_\pm^2)^{3/2}} + \OO(t^{-2}\ln |t|)\, .
\ee
Therefore the integral has the structure of $I_5$ and can be evaluated as
\be \label{ex1.10}
\tilde e^{(2)}_{ij}= i\, {m} \, \NN'\, \ln \omega^{-1}\, 
\left[ C_+ \, \beta_{+ i} \, \beta_{+ j} \, {1\over (1-\vec\beta_+^2)^{3/2}} -
C_- \, \beta_{- i} \, \beta_{- j} \, {1\over (1-\vec\beta_-^2)^{3/2}}\right]\, .
\ee

Adding \refb{ex1.9} and \refb{ex1.10} we get
\ben 
\tilde e_{ij} &=& \tilde e^{(1)}_{ij} + \tilde e^{(2)}_{ij}\nonumber \\ 
&=& -\,\omega^{-1} \, m\, \NN'\, 
\left\{ {1\over 1 - \hat n.\vec \beta_+} \, {1\over \sqrt{1-\vec \beta_+^2}}\, \beta_{+ i} 
\beta_{+ j} - {1\over 1 - \hat n.\vec \beta_-} \, {1\over \sqrt{1-\vec \beta_-^2}}\, \beta_{- i} 
\beta_{- j}\right\}\nonumber \\
&& - i\, m\, \NN'\,  \ln \omega^{-1} \left[ 
{1\over \sqrt{1-\vec \beta_+^2}}\, \beta_{+ i} 
\beta_{+ j} C_+ {1\over 1-\hat n.\vec\beta_+}  
- {1\over \sqrt{1-\vec \beta_-^2}}\, \beta_{- i} 
\beta_{- j} \, C_- {1\over 1-\hat n.\vec\beta_-} 
\right]\, .\nonumber \\
\een
Comparing this with \refb{ehabexp} we see that the soft graviton factor $\tilde S_{\rm gr}$,
extracted from classical radiation, is given by
\ben 
\tilde S_{\rm gr}(\ve, k)
&=& -\,\omega^{-1} \, m\, \ve^{ij}\, 
\left\{ {1\over 1 - \hat n.\vec \beta_+} \, {1\over \sqrt{1-\vec \beta_+^2}}\, \beta_{+ i} 
\beta_{+ j} - {1\over 1 - \hat n.\vec \beta_-} \, {1\over \sqrt{1-\vec \beta_-^2}}\, \beta_{- i} 
\beta_{- j}\right\}\nonumber \\
&& - i\, m\, \ve^{ij}\,  \ln \omega^{-1} \left[ 
{1\over \sqrt{1-\vec \beta_+^2}}\, \beta_{+ i} 
\beta_{+ j} C_+ {1\over 1-\hat n.\vec\beta_+}  
- {1\over \sqrt{1-\vec \beta_-^2}}\, \beta_{- i} 
\beta_{- j} \, C_- {1\over 1-\hat n.\vec\beta_-} 
\right]\, ,\nonumber \\
\een
for transverse polarization tensor $\ve$.
This agrees with $S_{\rm gr}$ given in \refb{egrgiven} 
upon replacing $\ln |t|$ by $\ln\omega^{-1}$. 

\subsection{Gravitational radiation from scattering via gravitational interaction}

We shall now consider the scattering of a probe of mass $m$ 
by a massive scatterer of mass $M_0$ due to gravitational 
interaction. We shall assume that the impact parameter (the distance of closest approach) is
large compared to the Schwarzschild  radius of the scatterer and work to first order in the ratio
of the Schwarzschild radius $M_0/(4\pi)$ 
and the impact parameter. The radiative part of the gravitational
field during such scattering was analyzed in \cite{peter}. After making appropriate changes in the signs and
normalization factors described in \cite{1801.07719}, 
it is given by a sum of four 
terms:
\be \label{esmgr}
\tilde e_{ij}= \tilde e^{(1)}_{ij}
+ \tilde e^{(2)}_{ij}+\tilde e^{(3)}_{ij}+\tilde e^{(4)}_{ij}\, .
\ee
$\tilde e^{(1)}$ is given by
\be \label{e25}
\tilde e^{(1)}_{ij}(\omega, \vec x) 
= {m\, e^{i\omega R} \over 4\, \pi\, R} \int {dt\over 1+2\vp(\vec r(t))} \, {dt\over d\sigma}\, v_i v_j \, 
e^{i\omega (t-
\hat n.\vec r(t))} +\hbox{boundary terms}
 \, ,
\ee
where $\vec r(t)$ denotes the trajectory of the particle,
\be\label{edefR'}
R\equiv |\vec x|, \qquad \hat n\equiv {\vec x\over |\vec x|}\, ,
\ee
and $\vp(\vec r)$ is the gravitational potential:
\be 
\vp(\vec r) = -{M_0\over 8\pi |\vec r|}\, ,
\ee
in the $8\pi G=1$ unit. The other $\tilde e^{(i)}$'s are given by
\ben \label{e26}
\tilde e^{(2)}_{ij}(\omega, \vec x) &=& i\, {M_0 m\over 32\, \pi^2\omega} {e^{i\omega R}\over R}
\int dt \, {dt\over d\sigma}\, (1+\vec v^2) \, \left(\p'_i\p'_j -{1\over 2} \delta_{ij} \, \p'_k\p'_k\right)\, 
\bigg\{ \ln {|\vec r^{\, \prime}|+\hat n.\vec r^{\, \prime}\over R} \, e^{i\omega (t - \hat n.\vec r^{\, \prime})} \nonumber \\ &&
\hskip 1in+ \int_{|\vec r^{\, \prime}|+\hat n.\vec r^{\, \prime}}^\infty {du\over u}
e^{i\omega (t - \hat n.\vec r^{\, \prime}+u)} 
\bigg\}\bigg|_{\vec r^{\, \prime}=\vec r(t)}\, , \quad \p'_i \equiv {\p\over \p r^{\prime i}}\, ,
\een
\ben  \label{e28}
\tilde e^{(3)}_{ij}(\omega, \vec x) &=& -i\, {M_0 m\over 16\, \pi^2} \, \omega \, {e^{i\omega R}\over R}
\int dt \, {dt\over d\sigma}\, v_i\, v_j \, 
\bigg\{ \ln {|\vec r(t)|
+\hat n.\vec r(t)\over R} \, e^{i\omega (t - \hat n.\vec r(t))} \nonumber \\ &&
\hskip 1in+ \int_{|\vec r(t)|+\hat n.\vec r(t)}^\infty {du\over u}
e^{i\omega (t - \hat n.\vec r(t)+u)} 
\bigg\}\, ,
\een
and
\ben \label{e28aa}
\tilde e^{(4)}_{ij}(\omega, \vec x) &=& - {M_0 m\over 16\, \pi^2} {e^{i\omega R}\over R}
\int dt \, {dt\over d\sigma}\, \left(v_i\p'_j + v_j \p'_i \right)\, 
\bigg\{ \ln {|\vec r^{\, \prime}|+\hat n.\vec r^{\, \prime}\over R} \, e^{i\omega (t - \hat n.\vec r^{\, \prime})} \nonumber \\ &&
\hskip 1in+ \int_{|\vec r^{\, \prime}|+\hat n.\vec r^{\, \prime}}^\infty {du\over u}
e^{i\omega (t - \hat n.\vec r^{\, \prime}+u)} 
\bigg\}\bigg|_{\vec r^{\, \prime}=\vec r(t)}\, ,
\een
where
\ben \label{edefdtds}
{dt\over d\sigma} &=& \left\{\left(1- {M_0\over 4\pi |\vec r(t)|}\right) - 
\left(1- {M_0\over 4\pi |\vec r(t)|}\right)^{-1} \vec v(t)^2
\right\}^{-1/2} \nonumber \\ 
&\simeq& 
{1\over \sqrt{1-\vec v(t)^2}} \, \left\{ 1 + {M_0\over 8\, \pi |\vec r(t)|} {1+\vec v(t)^2 \over 1-\vec v(t)^2}
\right\} \quad \hbox{for large $|\vec r(t)|$}\, .
\een

We begin with the evaluation of $\tilde e^{(1)}_{ij}$.
We have
\be \label{e534aa}
e^{i\omega(t- \hat n. \vec r(t))} =
{1\over i\omega} {1\over 1 - \hat n.\vec v(t)} {d\over dt}e^{i\omega(t - \hat n. \vec r(t))} \, .
\ee
Substituting this into \refb{e25} and integrating by parts we get
\be\label{e29}
\tilde e^{(1)}_{ij}(\omega, \vec x)  = -{m\over 4\pi\, R} e^{i\omega R} {1\over i\omega} 
\int dt \, e^{i\omega(t - \hat n. \vec r(t))} 
\, {d\over dt}\left[{1\over 1 - \hat n.\vec v(t)} {1\over 1+2\vp(\vec r(t))} \, {dt\over d\sigma}\, v_i v_j \right]\, .
\ee
Parametrizing $\vec r(t)$ for large $|t|$
as in \refb{e20a} and using \refb{edefdtds}
we get, as $t\to\pm\infty$, 
\ben\label{e30}
&& {1\over 1 - \hat n.\vec v(t)} {1\over 1+2\vp(\vec r(t))} \, {dt\over d\sigma}\, v_i v_j 
= {1\over 1 - \hat n.\vec v(t)} {1\over 1-M_0/(4\pi |\vec r(t)|)} \, {dt\over d\sigma} \, v_i v_j
\nonumber \\ &=& {1\over 1 - \hat n.\vec \beta_\pm} \, {1\over \sqrt{1-\vec \beta_\pm^2}}\, \beta_{\pm i} 
\beta_{\pm j}
\left[ 1  - {1\over t} \left\{C_\pm {1\over 1-\hat n.\vec\beta_\pm} \mp 
{ M_0\over 8\, \pi\, |\vec \beta_\pm|}\, {3-\vec\beta_\pm^2\over 1-\vec\beta_\pm^2} 
+C_\pm {1 \over 1 -\vec\beta_\pm^2} 
\right\}\right]\, , \nonumber \\ &&
(t - \hat n. \vec r(t))= t(1 -\hat n.\vec \beta_\pm) + C_\pm \hat n.\vec \beta_\pm \, \ln |t|\, .
\een
Comparing \refb{e29} with \refb{e2} and \refb{e30} with \refb{e1} we see that \refb{e29} takes the
form of the integral $I_1$ with
\ben
f_\pm &=& i\, {m\over 4\pi\, R} e^{i\omega R}  
{1\over 1 - \hat n.\vec \beta_\pm} \, {1\over \sqrt{1-\vec \beta_\pm^2}}\, \beta_{\pm i} 
\beta_{\pm j},\nonumber \\
\quad k_\pm &=& - i\, {m\over 4\pi\, R} e^{i\omega R}  
{1\over 1 - \hat n.\vec \beta_\pm} \, {1\over \sqrt{1-\vec \beta_\pm^2}}\, \beta_{\pm i} 
\beta_{\pm j} \nonumber \\ &&
\hskip 1in \left\{C_\pm {1\over 1-\hat n.\vec\beta_\pm} \mp 
{ M_0\over 8\, \pi\, |\vec \beta_\pm|}\, {3-\vec\beta_\pm^2\over 1-\vec\beta_\pm^2} 
 +C_\pm {1\over 1 -\vec\beta_\pm^2} 
\right\}, \nonumber \\ 
a_\pm &=& - (1 -\hat n.\vec \beta_\pm) \, .
\een
Therefore \refb{e2} gives
\ben \label{e39a}
\tilde e^{(1)}_{ij} &=& i\,\omega^{-1} {m\over 4\pi\, R} e^{i\omega R} 
\left\{ {1\over 1 - \hat n.\vec \beta_+} \, {1\over \sqrt{1-\vec \beta_+^2}}\, \beta_{+ i} 
\beta_{+ j} - {1\over 1 - \hat n.\vec \beta_-} \, {1\over \sqrt{1-\vec \beta_-^2}}\, \beta_{- i} 
\beta_{- j}\right\}\nonumber \\
&& - {m\over 4\pi\, R} e^{i\omega R}  \ln \omega^{-1} \left[ 
{1\over \sqrt{1-\vec \beta_+^2}}\, \beta_{+ i} 
\beta_{+ j} \left\{C_+ {1\over 1-\hat n.\vec\beta_+} - 
{ M_0\over 8\, \pi\, |\vec\beta_+|} \, {3-\vec\beta_+^2\over 1-\vec\beta_+^2}
+C_+ {1\over 1 -\vec\beta_+^2} 
\right\} \right.\nonumber \\ &&
\left. - {1\over \sqrt{1-\vec \beta_-^2}}\, \beta_{- i} 
\beta_{- j} \left\{C_- {1\over 1-\hat n.\vec\beta_-} + 
{M_0\over 8\pi |\vec\beta_-|} \, {3-\vec\beta_-^2\over 1-\vec\beta_-^2}
+C_- {1\over 1 -\vec\beta_-^2} 
\right\}
\right]\, .
\een

Next we turn to $\tilde e^{(3)}_{ij}$ given in \refb{e28}. Using \refb{e534aa}
and doing an integration by parts, we can express $\tilde e^{(3)}_{ij}$ as
\ben  
\tilde e^{(3)}_{ij}(\omega, \vec x) &=& {M_0 m\over 16\, \pi^2}  \, {e^{i\omega R}\over R}
\int dt \,   e^{i\omega(t-\hat n.\vec r(t))} {d\over dt} 
\left[ {1\over 1 -\hat n.\vec v(t)} \, {dt\over d\sigma}\, v_i\, v_j \right. \nonumber \\ &&
\left. \hskip 1in\bigg\{ \ln (|\vec r(t)|+\hat n.\vec r(t))  
+ \int_{|\vec r(t)|+\hat n.\vec r(t)}^\infty {du\over u}
e^{i\omega u} 
\bigg\}\right]\, .
\een
This integral is of the form $I_2$ given in \refb{e2} and therefore gives the result:\footnote{This term was
ignored in \cite{peter} since  at large impact parameter $\vec\beta_+\simeq \vec\beta_-$, and
the term inside the curly bracket of \refb{ee2fin} 
is small. However we can easily conceive a slightly different situation where a pair of particles undergo an
elastic collision in the black hole background, causing a change of order unity in each of their velocities.
In this case $\vec\beta_+-\vec\beta_-$ will be of order unity for each of these particles. The gravitational field
produced during this process will be given by the sum of the contributions due to these two particles, each of
which can be evaluated using the result given in this section.
}
\be\label{ee2fin}
\tilde e^{(3)}_{ij}=
- {M_0 m\over 16\, \pi^2}  \, \ln(\omega R) \, {e^{i\omega R}\over R} \, 
\left\{ {1\over 1 - \hat n.\vec \beta_+} \, {1\over \sqrt{1-\vec \beta_+^2}}\, \beta_{+ i} 
\beta_{+ j} - {1\over 1 - \hat n.\vec \beta_-} \, {1\over \sqrt{1-\vec \beta_-^2}}\, \beta_{- i} 
\beta_{- j}\right\}\, .
\ee
This term can be understood as arising from multiplication of the first line of \refb{e39a} 
by the phase factor $\exp[i \omega \, M_0\, \ln(\omega \, R) / (4\pi)]$. This
is precisely the additional phase factor \refb{ephase}
arising due to gravitational drag and backscattering experienced by the emitted radiation due to the
gravitational field of the mass $M_0$.

Next we consider $\tilde e^{(4)}_{ij}$ given in \refb{e28aa}. It can be expressed as
\ben 
\tilde e^{(4)}_{ij}(\omega, \vec x) &=& - {M_0 m\over 16\, \pi^2} {e^{i\omega R}\over R}
\int dt \, {dt\over d\sigma}\, \, e^{i\omega (t - \hat n.\vec r(t))} \, 
\nonumber \\
&& \left[ (-i\omega) \, \left(v_i n_j + v_j n_i \right) \, 
\bigg\{ \ln (|\vec r^{\, \prime}|+\hat n.\vec r^{\, \prime}) + \int_{|\vec r^{\, \prime}|
+\hat n.\vec r^{\, \prime}}^\infty {du\over u}
e^{i\omega u} 
\bigg\} \right. \nonumber \\ && \hskip -1in
\left. + {1\over |\vec r^{\, \prime}|+\hat n.\vec r^{\, \prime}} \left\{ v_i \left({r'_j\over |\vec r^{\, \prime}|} 
+ \hat n_j\right)
+ v_j \left({r'_i\over |\vec r^{\, \prime}|} + \hat n_i\right)\right\} \left\{ 1 - e^{i\omega(|\vec r^{\, \prime}|
+\hat n.\vec r^{\, \prime})}\right\}
\right]\Bigg|_{\vec r^{\, \prime}=\vec r(t)}\, .
\een
The contribution to the integral from the term in the second line vanishes after $\tilde e^{(4)}_{ij}$ is contracted with the polarization tensor
$\ve^{ij}$, since
$\ve_{ij}\hat n^j=\ve_{ij} k^j/|\vec k|=0$. 
The contribution from the last line has the same structure as $I_3$ and therefore
also does not generate any term proportional to $\omega^{-1}$
or $\ln \omega^{-1}$. 

We now turn to the computation of $\tilde e^{(2)}_{ij}$ given in \refb{e26}. With the gauge condition 
$\ve^i_{~i}=0$ given in \refb{epolcon}, 
the term proportional to $\delta_{ij}$ does not contribute to $\ve^{ij}\tilde e_{ij}$. Now
we have
\ben\label{e39}
&& \p'_j \bigg\{ \ln (|\vec r^{\, \prime}|+\hat n.\vec r^{\, \prime}) \, e^{i\omega (t - \hat n.\vec r^{\, \prime})} 
+ \int_{|\vec r^{\, \prime}|+\hat n.\vec r^{\, \prime}}^\infty {du\over u}
e^{i\omega (t - \hat n.\vec r^{\, \prime}+u)} 
\bigg\}\nonumber \\ 
&=& - i\, \omega\, \hat n_j \bigg\{ \ln (|\vec r^{\, \prime}|+\hat n.\vec r^{\, \prime}) \, e^{i\omega (t - \hat n.\vec r^{\, \prime})} 
+ \int_{|\vec r^{\, \prime}|+\hat n.\vec r^{\, \prime}}^\infty {du\over u}
e^{i\omega (t - \hat n.\vec r^{\, \prime}+u)} 
\bigg\} \nonumber \\ &&
+ {1\over |\vec r^{\, \prime}|+\hat n.\vec r^{\, \prime}}
\left({r'_j\over |\vec r^{\, \prime}|} + \hat n_j\right) \, \left\{ e^{i\omega (t - \hat n.\vec r^{\, \prime})} 
- e^{i\omega (t +|\vec r^{\, \prime}|)}
\right\}\, .
\een
Substituting this into \refb{e26} we see that the contribution from the term in the first line of the
right hand side of \refb{e39} will vanish after contraction with $\ve^{ij}$. This allows us
to focus on the term in the last line of \refb{e39}. Now we have 
\ben 
&& \p'_i \left[ {1\over |\vec r^{\, \prime}|+\hat n.\vec r^{\, \prime}}\left({r'_j\over |\vec r^{\, \prime}|} + \hat n_j\right) \, \left\{ e^{i\omega (t - \hat n.\vec r^{\, \prime})} 
- e^{i\omega (t +|\vec r^{\, \prime}|)}
\right\}
\right] \nonumber \\
&=& -{1\over (|\vec r^{\, \prime}|+\hat n.\vec r^{\, \prime})^2} \left({r'_i\over |\vec r^{\, \prime}|} + \hat n_i\right) \, 
\left({r'_j\over |\vec r^{\, \prime}|} + \hat n_j\right) \, \left\{ e^{i\omega (t - \hat n.\vec r^{\, \prime})} 
- e^{i\omega (t +|\vec r^{\, \prime}|)}
\right\}
\nonumber \\ && +{1\over |\vec r^{\, \prime}|+\hat n.\vec r^{\, \prime}}\left\{ {\delta_{ij}\over |\vec r^{\, \prime}|} - {r'_i r'_j\over (|\vec r^{\, \prime}|)^3}\right\}\, \left\{ e^{i\omega (t - \hat n.\vec r^{\, \prime})} 
- e^{i\omega (t +|\vec r^{\, \prime}|)}
\right\}\nonumber \\ &&
- i\, \omega\, {1\over |\vec r^{\, \prime}|+\hat n.\vec r^{\, \prime}}\left({r'_j\over |\vec r^{\, \prime}|} + \hat n_j\right) \, 
\left\{ e^{i\omega (t - \hat n.\vec r^{\, \prime})} \, \hat n_i
+ e^{i\omega (t +|\vec r^{\, \prime}|)} \, {r'_i\over |\vec r^{\, \prime}|}\right\}\, .
\een
Before substituting this into \refb{e26} we note that the term proportional to $\delta_{ij}$ does not
contribute to $\ve^{ij}\tilde e_{ij}$ due to the $\ve^i_{~i}=0$ condition. Also the terms proportional to
$\hat n_i$ and $\hat n_j$ can be dropped since $\hat n_i = k_i/|\vec k|$ and we have the 
$k_i \ve^{ij}=0$ condition in \refb{epolcon}. 
Substituting this into \refb{e26} we see that the relevant part of $\tilde e^{(2)}_{ij}$ is given by
\ben
\tilde e^{(2)}_{ij}(\omega, \vec x) 
&=& i\, {M_0 m\over 32\, \pi^2\omega} {e^{i\omega R}\over R}
\int dt \, {dt\over d\sigma}\, (1+\vec v^2) \nonumber \\ &&
\Bigg[-{1\over (|\vec r^{\, \prime}|+\hat n.\vec r^{\, \prime})^2}  {r'_i r'_j\over (|\vec r^{\, \prime}|)^2}  \, 
\left\{ e^{i\omega (t - \hat n.\vec r^{\, \prime})} 
- e^{i\omega (t +|\vec r^{\, \prime}|)}
\right\}
\nonumber \\ && 
-{1\over |\vec r^{\, \prime}|(|\vec r^{\, \prime}|+\hat n.\vec r^{\, \prime})}\, 
{r'_i r'_j\over (|\vec r^{\, \prime}|)^2}\, \left\{ e^{i\omega (t - \hat n.\vec r^{\, \prime})} 
- e^{i\omega (t +|\vec r^{\, \prime}|)}
\right\}\nonumber \\ &&
- i\, \omega\, {1\over |\vec r^{\, \prime}|+\hat n.\vec r^{\, \prime}}\, {r'_i \, r'_j\over (|\vec r^{\, \prime}|)^2}  \, 
 e^{i\omega (t +|\vec r^{\, \prime}|)} \Bigg]_{\vec r^{\, \prime} =\vec r(t)}
\, .
\een
Using the asymptotic behavior \refb{e20a} we see that the contribution from the second and third
line have the form $I_4$ and the contribution from the last line has the form $I_5$,
both given in \refb{e2}. The result
is
\ben \label{e47}
&& \tilde e^{(2)}_{ij}(\omega, \vec x) \nonumber \\ &=&\ln\omega^{-1}\, 
{M_0 m\over 32\, \pi^2} {e^{i\omega R}\over R}\, \left[ -{(1+\vec\beta^2) \beta_i\beta_j\over
\vec\beta^2 
\, \sqrt{1-\vec\beta^2}} \left\{{1\over (\eps|\vec\beta|+\hat n.\vec\beta)^2} + {1\over (\eps|\vec\beta|+\hat n.\vec\beta)
\eps|\vec\beta|}\right\} (\eps|\vec\beta|+\hat n.\vec\beta)
\right. \nonumber \\ &&\hskip 1.5in \left. + {(1+\vec\beta^2) \beta_i\beta_j\over
\vec\beta^2 \, \sqrt{1-\vec\beta^2}} {1\over (\eps|\vec\beta|+\hat n.\vec\beta)}
\right]_-^+\nonumber \\
&=& -\ln\omega^{-1}\, 
{M_0 m\over 32\, \pi^2} {e^{i\omega R}\over R}\, \left[{(1+\vec\beta^2) \beta_i\beta_j\over
\eps|\vec\beta|^3\, \sqrt{1-\vec\beta^2}}\right]_-^+ \, ,
\een
where $\eps$ is +1 for outgoing states and $-1$ for ingoing states. This gives
\be \label{e5.45e2}
\tilde e^{(2)}_{ij}(\omega, \vec x) =-\ln\omega^{-1}\, 
{M_0 m\over 32\, \pi^2} {e^{i\omega R}\over R}\, \left[{(1+\vec\beta_+^2) \beta_{+i}\beta_{+j}\over
|\vec\beta_+|^3\, \sqrt{1-\vec\beta_+^2}} + 
{(1+\vec\beta_ -^2) \beta_{ -i}\beta_{ -j}\over
|\vec\beta_ -|^3\, \sqrt{1-\vec\beta_ -^2}}\right]\, .
\ee

Adding \refb{e39a}, \refb{ee2fin} and \refb{e5.45e2}, 
using \refb{esmgr} and comparing the result with \refb{ehabexp} 
with the extra phase factor \refb{ephase} on the right hand side (which cancels \refb{ee2fin}),
we get the following prediction for the soft factor from the classical scattering results:
\ben \label{e5.46}
\tilde S_{\rm gr} &=& -m\, \omega^{-1}\, \ve^{ij}\, 
\left\{ {1\over 1 - \hat n.\vec \beta_+} \, {1\over \sqrt{1-\vec \beta_+^2}}\, \beta_{+ i} 
\beta_{+ j} - {1\over 1 - \hat n.\vec \beta_-} \, {1\over \sqrt{1-\vec \beta_-^2}}\, \beta_{- i} 
\beta_{- j}\right\}\nonumber \\
&& \hskip -.5in - i\, {m} \,   \ln \omega^{-1} \, \ve^{ij}\, \left[ 
{1\over \sqrt{1-\vec \beta_+^2}}\, \beta_{+ i} 
\beta_{+ j} \left\{C_+ {1\over 1-\hat n.\vec\beta_+} -
{ M_0\over 8\, \pi\, |\vec\beta_+|^3}\, {3\vec\beta_+^2 -1 \over 1-\vec \beta_+^2}
+C_+ {1\over 1 -\vec\beta_+^2} 
\right\} \right.\nonumber \\ &&
\left. - {1\over \sqrt{1-\vec \beta_-^2}}\, \beta_{- i} 
\beta_{- j} \left\{C_- {1\over 1-\hat n.\vec\beta_-} + 
{M_0\over 8\, \pi\, |\vec\beta_-|} {3\vec\beta_-^2-1\over 1-\vec\beta_-^2}
+C_- {1\over 1 -\vec\beta_-^2} 
\right\}
\right]\, .
\een

In order to compare this to \refb{egrgiven} we need to find the relation between $M_0$ and $C_\pm$.
Let $\vec v(t)$ be the velocity of the particle at large $|t|$ when the particle is at a distance $r$
from the black hole and $\vec\beta$ be the velocity as $|t|\to\infty$.
The expression of the total energy of the particle  in the $8\pi G=1$ units is given by
\be
E = m\left(1- {M_0\over 4\, \pi\, r}\right) \left\{ \left(1- {M_0\over 4\, \pi\, r}\right) - 
\left(1- {M_0\over 4\, \pi\, r}\right)^{-1} \vec v^2\right\}^{-1/2} \, ,
\ee
so that the conservation of energy gives
\be
\left(1- {M_0\over 4\, \pi\, r}\right) \left\{ \left(1- {M_0\over 4\, \pi\, r}\right) - 
\left(1- {M_0\over 4\, \pi\, r}\right)^{-1} \vec v^2\right\}^{-1/2}
= (1-\vec \beta^2)^{-1/2}\, .
\ee
To first order in an expansion in powers of $M_0$ this gives
\be
\vec v(t) = \vec\beta \left( 1 + {M_0\over 8\pi \, \vec\beta^2 \, r} (1-3\vec\beta^2)\right)
=  \vec\beta \left( 1 + {M_0\over 8\pi \, |\vec\beta|^3 \, |t|} (1-3\vec\beta^2)\right)
\ee
where we have used $r=|\vec\beta||t|$. Comparing this with \refb{e20a} we get
\be 
C_\pm = \mp  {M_0 (1-3\vec\beta_\pm^2)\over 8\pi |\vec \beta_\pm|^3} \, .
\ee
Using this we can express \refb{e5.46} as
\ben \label{e5.46new}
\tilde S_{\rm gr} &=& -m\, \omega^{-1}\, \ve^{ij}\, 
\left\{ {1\over 1 - \hat n.\vec \beta_+} \, {1\over \sqrt{1-\vec \beta_+^2}}\, \beta_{+ i} 
\beta_{+ j} - {1\over 1 - \hat n.\vec \beta_-} \, {1\over \sqrt{1-\vec \beta_-^2}}\, \beta_{- i} 
\beta_{- j}\right\}\nonumber \\
&& \hskip -1in - i\, {m} \, \ve^{ij}\,   \ln \omega^{-1} \left[ 
{1\over \sqrt{1-\vec \beta_+^2}}\, \beta_{+ i} 
\beta_{+ j} \, C_+ {1\over 1-\hat n.\vec\beta_+} - {1\over \sqrt{1-\vec \beta_-^2}}\, \beta_{- i} 
\beta_{- j} \, C_- {1\over 1-\hat n.\vec\beta_-} \right]\, . 
\een
This agrees with \refb{egrgiven} with $\ln |t|$ replaced by $\ln\omega^{-1}$.

\bigskip

{\bf Acknowledgement:} 
We wish to thank Miguel Campiglia, Walter Golberger, Ira Rothstein and
Biswajit Sahoo for discussions. 
Work of A.L. is supported  in part by Ramanujan Fellowship. 
The work of A.S. was
supported in part by the 
J. C. Bose fellowship of 
the Department of Science and Technology, India.

\appendix

\sectiono{Evaluation of some integrals} \label{sa}

Our goal in this appendix will be to compute the $1/\omega$ and $\ln\omega$ terms in 
the following integrals in $\omega\to 0$ limit. 
\be
I_1\equiv {1\over \omega} \int_{-\infty}^\infty dt \, e^{-i\,\omega\, g(t) } f'(t) \, ,
\ee
\be
I_2\equiv  \int_{-\infty}^\infty dt \,e^{-i\,\omega\, g(t) }  {d\over dt} \left[f(t) \left\{
\ln {h(t)\over R}+ \int_{h(t)}^\infty  e^{i\, \omega\, u} \, {du\over u}\right\}\right] \, ,
\ee
\be I_3\equiv \int_{-\infty}^\infty dt \, {1\over r(t)} \, f(t)\, \left[e^{-i\, \omega\, g(t)}-
e^{-i\, \omega\, h(t)}\right] \, ,
\ee
\be I_4\equiv {1\over \omega} 
\int_{-\infty}^\infty dt \, {1\over r(t)^2} \, f(t)\, \left[e^{-i\, \omega\, g(t)}-
e^{-i\, \omega\, h(t)}\right]\, ,
\ee
\be
I_5 = \int_{-\infty}^\infty dt \, {1\over r(t)} \, f(t)\, e^{-i\, \omega\, g(t)}\, ,
\ee
where, as described in \refb{e1}, 
$f(t), g(t), h(t)$ are smooth functions with the property 
\ben \label{e1app}
&&  f(t) = f_\pm+{k_\pm\over t}, \quad 
g(t) \to a_\pm t + b_\pm \ln |t|, \nonumber \\ &&
h(t) \to p_\pm t + q_\pm \ln |t|, \quad r(t)\to c_\pm \, t+ d_\pm \ln |t|, 
\quad \hbox{as} \quad t\to\pm\infty\, .
\een
We shall evaluate the integrals by separately estimating their contributions from the four regions:
$|t|\sim 1$, $1<<|t<<\omega^{-1}$, $|t|\sim\omega^{-1}$ and $|t|>>\omega^{-1}$.

\subsection{Evaluation of $I_1$}

We express $I_1$ as
\be
I_1 = {1\over \omega} \int_{-\infty}^\infty dt \,  f'(t) 
+ {1\over \omega} \int_{-\infty}^\infty dt \, \left\{e^{-i\,\omega\, g(t) }-1\right\} f'(t) \, .
\ee
The first term gives $\omega^{-1} (f_+-f_-)$. The second term can be evaluated by dividing the integration
region into different segments. In the region $t\sim 1$ the term inside the curly bracket 
is of order $\omega$ and we get a
finite contribution. In the region $1<<|t|<<\omega^{-1}$ we can approximate the integral as
\be \label{eexright}
-i \int_{1<<|t|<<\omega^{-1}} dt   \, g(t) \, f'(t) \simeq i \int_{1<<|t|<<\omega^{-1}} dt   \, {a_\pm k_\pm \over t}
\simeq  i\,(a_+k_+ - a_- k_-) \ln \omega^{-1}\, .
\ee
The last step can be justified
as follows. Let us fix the integration range to be 
$[a,b\, \omega^{-1}]$ where $a$ and $b$ some fixed
numbers with $a\ >>\ 1$ 
and $b\ <<1$.  The right hand side of the above equation can then be approximated as 
\be
i\,(a_+k_+ - a_- k_-) \left[ \ln\omega^{-1} + \ln\frac{b}{a}\right]\, .
\ee
Even though $\frac{b}{a}\ <<1$, as $\omega^{-1}$ becomes large, we can 
ignore $\ln \frac{b}{a}$ compared to $\ln\omega^{-1}$, arriving at the right hand side of \refb{eexright}.

In the region $|t| \sim \omega^{-1}$ and $|t|>\omega^{-1}$ the magnitude of the
integral is bounded by a term of order 
\be
\omega^{-1} \, \int_{|t|>\omega^{-1}} \, 2\, \{|k_\pm| \, t^{-2}\}  \, dt \sim  2\, |k_\pm| \, .
\ee
Therefore for small $\omega$, $I_1$ can be estimated to be
\be\label{e9}
I_1 = \omega^{-1} (f_+-f_-) + i\, (a_+k_+ - a_- k_-) \ln \omega^{-1} + \hbox{finite}\, .
\ee

\subsection{Evaluation of $I_2$}
 
Let us express $I_2$ as
\be
I_2\equiv  \int_{-\infty}^\infty dt \,e^{-i\,\omega\, g(t) }  \left[f'(t) \left\{
\ln {h(t)\over R}+ \int_{h(t)}^\infty  e^{i\, \omega\, u} \, {du\over u}\right\}
+ f(t) \, h'(t)\, h(t)^{-1}\,  \left(1 - e^{i\omega h(t)}
\right)\right] \, .
\ee
While integrating over the region $|t|\sim 1$, we can replace $e^{i\omega t}$ by 1. 
Also in this region the integral inside the curly bracket can be evaluated by changing variable
from $u$ to $v=\omega\, u$, and yields $\ln\omega^{-1}$ plus a finite term. Therefore the
term inside the curly bracket is given by $-\ln (R\, \omega)$ plus a finite
term, and the integration over $t$ produces a term
\be 
-\ln(R\, \omega) (f_+-f_-) + \hbox{finite}\, .
\ee
For $1<< |t|<<\omega^{-1}$ the term with the $f'(t)$ factor is of order $t^{-2}\times$
logarithmic terms and produces a 
finite result. On the other hand the $f(t) h'(t) (h(t))^{-1}(1-e^{i\omega h(t)})$ factor is of order $-i\,\omega \,
f_\pm p_\pm$ and gives negligible contribution to the integral from the $1<<|t|<<\omega^{-1}$ region.
For $|t|\sim \omega^{-1}$ the integrand is of order $t^{-1}$ and therefore gives a finite contribution to the
integral. Finally for $t>>\omega^{-1}$ the term proportional to $f'(t)$ falls off as $t^{-2}\times$
logarithmic terms and its contribution
to the integral is vanishes in the $\omega\to 0$ limit. In this range 
the  term proportional to $f(t) \, h'(t)\, (h(t))^{-1}$ may be
approximated as
\be\label{ea.12}
\int_{\omega^{-1}}^\infty \, dt \, f_\pm \, t^{-1} \, \left\{ e^{-i a_\pm \omega t} - e^{i(1-a_\pm)\omega t}\right\}\, .
\ee
After changing variable to $u=-a_\pm \omega t$ in the first term and $ (1-a_\pm)\omega t$ in the second term,
each of the integrals can be converted to the form
\be
f_\pm \, \int^\infty du \, u^{-1} \, e^{iu}\, ,
\ee
with finite lower limit of order unity. This gives a finite result.
Therefore we get 
\be 
I_2=-\ln(R\, \omega) \, (f_+-f_-) + \hbox{finite}\, .
\ee

\subsection{Evaluation of $I_3\equiv \int_{-\infty}^\infty dt \, (r(t))^{-1} \, f(t)\, \left[e^{-i\, \omega\, g(t)}-
e^{-i\, \omega\, h(t)}\right]$}

The region $|t|\sim 1$ gives a finite contribution. 
For $1<< |t|<<\omega^{-1}$ the integrand may be approximated as
\be (c_\pm)^{-1} \, f_\pm \, (-i\omega) \, (a_\pm -p_\pm) \, ,
\ee
and the integral receives negligible contribution from this region.

For $|t|\sim \omega^{-1}$ the term in the square bracket is of order unity. 
But the rest of the integrand is of order
$ (c_\pm\, t)^{-1} \, f_\pm$ and integration over $t$ in the range 
$|t|\sim \omega^{-1}$ produces at most a term
of order unity -- there is no contribution proportional to $\ln\,\omega$.
Finally for $|t|>>\omega^{-1}$
the integrand has the same form as \refb{ea.12} with $(1-a_\pm)$ replaced by $p_\pm$ in the
second exponent, and an overall multiplicative factor $(c_\pm)^{-1}$.
Therefore it gives a finite result.

\subsection{Evaluation of $I_4\equiv  \omega^{-1}\, 
\int_{-\infty}^\infty dt \, (r(t))^{-2} \, f(t)\, \left[e^{-i\, \omega\, g(t)}-
e^{-i\, \omega\, h(t)}\right]$}

We use 
\be 
{1\over r(t)^2} = -{d\over dt} \left\{ {1\over r(t)}\right\} {1\over r'(t)}\, .
\ee
Substituting this into the expression for $I_4$ and doing an integration by parts  we get the following form
of the integral for large $|t|$:
\be
{1\over  \omega} \int dt \, {1\over r(t)} 
 \, {d\over dt} \left( {f(t)\over r'(t)}\right) \, \left[e^{-i\, \omega\, g(t)}-
e^{-i\, \omega\, h(t)}\right] 
-{i} \int dt \, {f(t)\over r(t) \, r'(t)} 
 \,   \left[g'(t) \, e^{-i\, \omega\, g(t)}- h'(t)\, 
e^{-i\, \omega\, h(t)}\right] \, .\ee
For $|t|\sim 1$ the integrands in both terms are finite in the $\omega\to 0$ limit and we get finite
contribution to the integral. Using \refb{e1app} we see that  in the first  
term, part of  the integrand outside the square bracket falls off as $1/|t|^3$ for $|t|>>1$. 
On the other hand, using the inequality $|\sin u|\le |u|$, we can see that the terms inside the square
bracket of the first  term is bounded by $\omega|g(t)-h(t)|\sim \omega\, |t|\, |a_\pm -p_\pm|$. 
 Therefore integration over the region $|t|>1$ yields a finite result as $\omega\to 0$ for the first term.

The contribution from the second term can be evaluated by noting that for large $|t|$, 
$f(t)/ \{ r(t)\, r'(t)\}\to f_\pm c_\pm^{-2} t^{-1}$, 
$g'(t)\to a_\pm$ and $h'(t)\to p_\pm$. Therefore in the $1<<|t|<<\omega^{-1}$ region the integrand
behaves as $-i \, f_\pm\, c_\pm^{-2} \, t^{-1} \, (a_\pm - p_\pm)$ and the dominant contribution
to the integral is given by
\be 
- i\, \{f_+ \, c_+^{-2}(a_+ - p_+) - f_-\, c_-^{-2}(a_- - p_-)\} \, \ln \, \omega^{-1} \, .
\ee
For $|t|\sim \omega^{-1}$ the integrand is of order $t^{-1}$, producing a finite result for the integral.
Finally for $|t|>>\omega^{-1}$ the integrand is proportional to $t^{-1} [g'(t) e^{-i\omega g(t)} -
h'(t) e^{-i\omega h(t)}]$. Each of these produces a finite contribution in the $\omega\to 0$ limit.

Therefore we get
\be 
I_4 \simeq - i \, \{f_+\, c_+^{-2}(a_+ - p_+) - f_-\, c_-^{-2}(a_- - p_-)\} \, \ln \, \omega^{-1} \, .
\ee

\subsection{Evaluation of $I_5\equiv \int_{-\infty}^\infty dt \, (r(t))^{-1} \, f(t)\, e^{-i\, \omega\, g(t)}$}

The $|t|\sim 1$  and $|t|\sim \omega^{-1}$
regions give finite contributions. The region  $1<<|t|<<\omega^{-1}$ gives
\be
\int_{1<<|t|<<\omega^{-1}} dt\, {f_\pm\over c_\pm t} \simeq \pm {f_\pm\over c_\pm} \, \ln \omega^{-1}\, .
\ee
Finally in the region $|t|>>\omega^{-1}$ the integral takes the form
\be
\int_{|t|>>\omega^{-1}} dt\, {f_\pm\over c_\pm t} e^{-ia_\pm \omega\, t -i b_\pm \omega \ln |t|}\, .
\ee
This is bounded by a finite number. Therefore the net contribution to $I_5$ is given by
\be
I_5\simeq (f_+ c_+^{-1} - f_- c_-^{-1}) \, \ln \omega^{-1}\, .
\ee
Note that the result for $I_3$ follows from this.


\begin{thebibliography}{99}



\small

\baselineskip=14pt

\parskip=0pt

\bibitem{Gell-Mann}
M.~Gell-Mann and M.~L.~Goldberger, Phys.\ Rev.\ {\bf 96}, 1433 (1954).

\bibitem{low}
F.~E.~Low, Phys.\ Rev.\ {\bf 110}, 974 (1958).

\bibitem{saito}
S.~Saito, Phys.\ Rev.\ {\bf 184}, 1894 (1969).

\bibitem{burnett}
T.~H.~Burnett and N.~M.~Kroll, Phys.\ Rev.\ Lett.\ {\bf 20}, 86 (1968).

\bibitem{bell}
J.~S.~Bell and R. Van Royen, Nuovo Cim.\ {\bf A60}, 62 (1969).

\bibitem{duca}
V.~Del Duca, Nucl. Phys. {\bf B345}, 369 (1990).


\bibitem{weinberg1} 
  S.~Weinberg,
  ``Photons and Gravitons in s Matrix Theory: 
  Derivation of Charge Conservation and Equality of Gravitational and Inertial Mass,''
  Phys.\ Rev.\  {\bf 135}, B1049 (1964).
  doi:10.1103/PhysRev.135.B1049

\bibitem{weinberg2} 
  S.~Weinberg,
  ``Infrared photons and gravitons,''
  Phys.\ Rev.\  {\bf 140}, B516 (1965).
  doi:10.1103/PhysRev.140.B516

\bibitem{jackiw1} 
  D.~J.~Gross and R.~Jackiw,
  ``Low-Energy Theorem for Graviton Scattering,''
  Phys.\ Rev.\  {\bf 166}, 1287 (1968).
  doi:10.1103/PhysRev.166.1287
  
\bibitem{jackiw2} 
  R.~Jackiw,
  ``Low-Energy Theorems for Massless Bosons: Photons and Gravitons,''
  Phys.\ Rev.\  {\bf 168}, 1623 (1968).
  doi:10.1103/PhysRev.168.1623
  
  
\bibitem{ademollo} 
  M.~Ademollo, A.~D'Adda, R.~D'Auria, F.~Gliozzi, E.~Napolitano, S.~Sciuto and P.~Di Vecchia,
  ``Soft Dilations and Scale Renormalization in Dual Theories,''
  Nucl.\ Phys.\ B {\bf 94}, 221 (1975).
  doi:10.1016/0550-3213(75)90491-5

\bibitem{shapiro} 
  J.~A.~Shapiro,
  ``On the Renormalization of Dual Models,''
  Phys.\ Rev.\ D {\bf 11}, 2937 (1975).
  doi:10.1103/PhysRevD.11.2937



\bibitem{1103.2981} 
  C.~D.~White,
  ``Factorization Properties of Soft Graviton Amplitudes,''
  JHEP {\bf 1105}, 060 (2011)
  doi:10.1007/JHEP05(2011)060
  [arXiv:1103.2981 [hep-th]].

\bibitem{1404.4091} 
  F.~Cachazo and A.~Strominger,
  ``Evidence for a New Soft Graviton Theorem,''
  arXiv:1404.4091 [hep-th].

\bibitem{1404.5551} 
  E.~Casali,
  ``Soft sub-leading divergences in Yang-Mills amplitudes,''
  JHEP {\bf 1408}, 077 (2014)
  doi:10.1007/JHEP08(2014)077
  [arXiv:1404.5551 [hep-th]].

\bibitem{1404.7749} 
  B.~U.~W.~Schwab and A.~Volovich,
  ``Subleading Soft Theorem in Arbitrary Dimensions from Scattering Equations,''
  Phys.\ Rev.\ Lett.\  {\bf 113}, no. 10, 101601 (2014)
  doi:10.1103/PhysRevLett.113.101601
  [arXiv:1404.7749 [hep-th]].

\bibitem{1405.1015} 
  Z.~Bern, S.~Davies and J.~Nohle,
  ``On Loop Corrections to Subleading Soft Behavior of Gluons and Gravitons,''
  Phys.\ Rev.\ D {\bf 90}, no. 8, 085015 (2014)
  doi:10.1103/PhysRevD.90.085015
  [arXiv:1405.1015 [hep-th]].

\bibitem{1405.1410} 
  S.~He, Y.~t.~Huang and C.~Wen,
  ``Loop Corrections to Soft Theorems in Gauge Theories and Gravity,''
  JHEP {\bf 1412}, 115 (2014)
  doi:10.1007/JHEP12(2014)115
  [arXiv:1405.1410 [hep-th]].

\bibitem{1405.2346} 
  A.~J.~Larkoski,
  ``Conformal Invariance of the Subleading Soft Theorem in Gauge Theory,''
  Phys.\ Rev.\ D {\bf 90}, no. 8, 087701 (2014)
  doi:10.1103/PhysRevD.90.087701
  [arXiv:1405.2346 [hep-th]].


\bibitem{1405.3413} 
  F.~Cachazo and E.~Y.~Yuan,
  ``Are Soft Theorems Renormalized?,''
  arXiv:1405.3413 [hep-th].

\bibitem{1405.3533} 
  N.~Afkhami-Jeddi,
  ``Soft Graviton Theorem in Arbitrary Dimensions,''
  arXiv:1405.3533 [hep-th].

\bibitem{1406.4172} 
  B.~U.~W.~Schwab,
  ``Subleading Soft Factor for String Disk Amplitudes,''
  JHEP {\bf 1408}, 062 (2014)
  doi:10.1007/JHEP08(2014)062
  [arXiv:1406.4172 [hep-th]].

\bibitem{1406.5155} 
  M.~Bianchi, S.~He, Y.~t.~Huang and C.~Wen,
  ``More on Soft Theorems: Trees, Loops and Strings,''
  Phys.\ Rev.\ D {\bf 92}, no. 6, 065022 (2015)
  doi:10.1103/PhysRevD.92.065022
  [arXiv:1406.5155 [hep-th]].

\bibitem{1406.6574} 
  J.~Broedel, M.~de Leeuw, J.~Plefka and M.~Rosso,
  ``Constraining subleading soft gluon and graviton theorems,''
  Phys.\ Rev.\ D {\bf 90}, no. 6, 065024 (2014)
  doi:10.1103/PhysRevD.90.065024
  [arXiv:1406.6574 [hep-th]].

\bibitem{1406.6987} 
  Z.~Bern, S.~Davies, P.~Di Vecchia and J.~Nohle,
  ``Low-Energy Behavior of Gluons and Gravitons from Gauge Invariance,''
  Phys.\ Rev.\ D {\bf 90}, no. 8, 084035 (2014)
  doi:10.1103/PhysRevD.90.084035
  [arXiv:1406.6987 [hep-th]].
  
\bibitem{1406.7184} 
  C.~D.~White,
  ``Diagrammatic insights into next-to-soft corrections,''
  Phys.\ Lett.\ B {\bf 737}, 216 (2014)
  doi:10.1016/j.physletb.2014.08.041
  [arXiv:1406.7184 [hep-th]].
  
\bibitem{1407.5936} 
  M.~Zlotnikov,
  ``Sub-sub-leading soft-graviton theorem in arbitrary dimension,''
  JHEP {\bf 1410}, 148 (2014)
  doi:10.1007/JHEP10(2014)148
  [arXiv:1407.5936 [hep-th]].

\bibitem{1407.5982} 
  C.~Kalousios and F.~Rojas,
  ``Next to subleading soft-graviton theorem in arbitrary dimensions,''
  JHEP {\bf 1501}, 107 (2015)
  doi:10.1007/JHEP01(2015)107
  [arXiv:1407.5982 [hep-th]].

\bibitem{1408.4179} 
  Y.~J.~Du, B.~Feng, C.~H.~Fu and Y.~Wang,
  ``Note on Soft Graviton theorem by KLT Relation,''
  JHEP {\bf 1411}, 090 (2014)
  doi:10.1007/JHEP11(2014)090
  [arXiv:1408.4179 [hep-th]].

\bibitem{1410.6406} 
  D.~Bonocore, E.~Laenen, L.~Magnea, L.~Vernazza and C.~D.~White,
  ``The method of regions and next-to-soft corrections in DrellÐYan production,''
  Phys.\ Lett.\ B {\bf 742}, 375 (2015)
  doi:10.1016/j.physletb.2015.02.008
  [arXiv:1410.6406 [hep-ph]].

\bibitem{1411.6661} 
  B.~U.~W.~Schwab,
  ``A Note on Soft Factors for Closed String Scattering,''
  JHEP {\bf 1503}, 140 (2015)
  doi:10.1007/JHEP03(2015)140
  [arXiv:1411.6661 [hep-th]].

\bibitem{1412.3699} 
  A.~Sabio Vera and M.~A.~Vazquez-Mozo,
  ``The Double Copy Structure of Soft Gravitons,''
  JHEP {\bf 1503}, 070 (2015)
  doi:10.1007/JHEP03(2015)070
  [arXiv:1412.3699 [hep-th]].

\bibitem{1502.05258}
 P.~Di Vecchia, R.~Marotta and M.~Mojaza,
  ``Soft theorem for the graviton, dilaton and the Kalb-Ramond field in the bosonic string,''
  JHEP {\bf 1505}, 137 (2015)
  doi:10.1007/JHEP05(2015)137
  [arXiv:1502.05258 [hep-th]].




\bibitem{1503.04816} 
  F.~Cachazo, S.~He and E.~Y.~Yuan,
  ``New Double Soft Emission Theorems,''
  Phys.\ Rev.\ D {\bf 92}, no. 6, 065030 (2015)
  doi:10.1103/PhysRevD.92.065030
  [arXiv:1503.04816 [hep-th]].

\bibitem{1504.01364} 
  A.~E.~Lipstein,
  ``Soft Theorems from Conformal Field Theory,''
  JHEP {\bf 1506}, 166 (2015)
  doi:10.1007/JHEP06(2015)166
  [arXiv:1504.01364 [hep-th]].
  
  
\bibitem{1504.05558} 
  T.~Klose, T.~McLoughlin, D.~Nandan, J.~Plefka and G.~Travaglini,
  ``Double-Soft Limits of Gluons and Gravitons,''
  JHEP {\bf 1507}, 135 (2015)
  doi:10.1007/JHEP07(2015)135
  [arXiv:1504.05558 [hep-th]].

\bibitem{1504.05559} 
  A.~Volovich, C.~Wen and M.~Zlotnikov,
  ``Double Soft Theorems in Gauge and String Theories,''
  JHEP {\bf 1507}, 095 (2015)
  doi:10.1007/JHEP07(2015)095
  [arXiv:1504.05559 [hep-th]].

\bibitem{1505.05854} 
  M.~Bianchi and A.~L.~Guerrieri,
  ``On the soft limit of open string disk amplitudes with massive states,''
  JHEP {\bf 1509}, 164 (2015)
  doi:10.1007/JHEP09(2015)164
  [arXiv:1505.05854 [hep-th]].

\bibitem{1507.00938} 
  P.~Di Vecchia, R.~Marotta and M.~Mojaza,
  ``Double-soft behavior for scalars and gluons from string theory,''
  JHEP {\bf 1512}, 150 (2015)
  doi:10.1007/JHEP12(2015)150
  [arXiv:1507.00938 [hep-th]].

\bibitem{1507.08829} 
  A.~L.~Guerrieri,
  ``Soft behavior of string amplitudes with external massive states,''
  Nuovo Cim.\ C {\bf 39}, no. 1, 221 (2016)
  doi:10.1393/ncc/i2016-16221-2
  [arXiv:1507.08829 [hep-th]].

\bibitem{1507.08882} 
  S.~D.~Alston, D.~C.~Dunbar and W.~B.~Perkins,
  ``$n$-point amplitudes with a single negative-helicity graviton,''
  Phys.\ Rev.\ D {\bf 92}, no. 6, 065024 (2015)
  doi:10.1103/PhysRevD.92.065024
  [arXiv:1507.08882 [hep-th]].

\bibitem{1509.07840} 
  Y.~t.~Huang and C.~Wen,
  ``Soft theorems from anomalous symmetries,''
  JHEP {\bf 1512}, 143 (2015)
  doi:10.1007/JHEP12(2015)143
  [arXiv:1509.07840 [hep-th]].

\bibitem{1511.04921} 
  P.~Di Vecchia, R.~Marotta and M.~Mojaza,
  ``Soft Theorems from String Theory,''
  Fortsch.\ Phys.\  {\bf 64}, 389 (2016)
  doi:10.1002/prop.201500068
  [arXiv:1511.04921 [hep-th]].
  
  \bibitem{1512.00803} 
  M.~Bianchi and A.~L.~Guerrieri,
  ``On the soft limit of closed string amplitudes with massive states,''
  Nucl.\ Phys.\ B {\bf 905}, 188 (2016)
  doi:10.1016/j.nuclphysb.2016.02.005
  [arXiv:1512.00803 [hep-th]].

\bibitem{1601.03457} 
  M.~Bianchi and A.~L.~Guerrieri,
  ``On the soft limit of tree-level string amplitudes,''
  arXiv:1601.03457 [hep-th].


\bibitem{1604.00650} 
  J.~Rao and B.~Feng,
  ``Note on Identities Inspired by New Soft Theorems,''
  JHEP {\bf 1604}, 173 (2016)
  doi:10.1007/JHEP04(2016)173
  [arXiv:1604.00650 [hep-th]].

\bibitem{1604.03355} 
  P.~Di Vecchia, R.~Marotta and M.~Mojaza,
  ``Subsubleading soft theorems of gravitons and dilatons in the bosonic string,''
  JHEP {\bf 1606}, 054 (2016)
  doi:10.1007/JHEP06(2016)054
  [arXiv:1604.03355 [hep-th]].
  
 
  \bibitem{1604.02834} 
  S.~He, Z.~Liu and J.~B.~Wu,
  ``Scattering Equations, Twistor-string Formulas and Double-soft Limits 
  in Four Dimensions,''
  JHEP {\bf 1607}, 060 (2016)
  doi:10.1007/JHEP07(2016)060
  [arXiv:1604.02834 [hep-th]].

\bibitem{1604.03893} 
  F.~Cachazo, P.~Cha and S.~Mizera,
  ``Extensions of Theories from Soft Limits,''
  JHEP {\bf 1606}, 170 (2016)
  doi:10.1007/JHEP06(2016)170
  [arXiv:1604.03893 [hep-th]].

\bibitem{1607.02700} 
  A.~P.~Saha,
  ``Double Soft Theorem for Perturbative Gravity,''
  JHEP {\bf 1609}, 165 (2016)
  doi:10.1007/JHEP09(2016)165
  [arXiv:1607.02700 [hep-th]].

\bibitem{1610.03481} 
  P.~Di Vecchia, R.~Marotta and M.~Mojaza,
  ``Soft behavior of a closed massless state in superstring and universality in the soft behavior of the dilaton,''
  JHEP {\bf 1612}, 020 (2016)
  doi:10.1007/JHEP12(2016)020
  [arXiv:1610.03481 [hep-th]].

\bibitem{1611.02172} 
  A.~Luna, S.~Melville, S.~G.~Naculich and C.~D.~White,
  ``Next-to-soft corrections to high energy scattering in QCD and gravity,''
  JHEP {\bf 1701}, 052 (2017)
  doi:10.1007/JHEP01(2017)052
  [arXiv:1611.02172 [hep-th]].

\bibitem{1611.07534} 
  H.~Elvang, C.~R.~T.~Jones and S.~G.~Naculich,
  ``Soft Photon and Graviton Theorems in Effective Field Theory,''
  arXiv:1611.07534 [hep-th].

\bibitem{1611.03137} 
  C.~Cheung, K.~Kampf, J.~Novotny, C.~H.~Shen and J.~Trnka,
  ``A Periodic Table of Effective Field Theories,''
  arXiv:1611.03137 [hep-th].



\bibitem{1702.02350} 
  A.~P.~Saha,
  ``Double Soft Theorem for Perturbative Gravity II: Some Details on CHY Soft Limits,''
  arXiv:1702.02350 [hep-th].

\bibitem{1702.03934} 
  A.~Sen,
  ``Soft Theorems in Superstring Theory,''
  arXiv:1702.03934 [hep-th].

\bibitem{1703.00024} 
  A.~Sen,
  ``Subleading Soft Graviton Theorem for Loop Amplitudes,''
  arXiv:1703.00024 [hep-th].


\bibitem{1705.06175} 
  P.~Di Vecchia, R.~Marotta and M.~Mojaza,
  ``Double-soft behavior of the dilaton of spontaneously broken conformal invariance,''
  arXiv:1705.06175 [hep-th].

\bibitem{1706.00759} 
  A.~Laddha and A.~Sen,
  ``Sub-subleading Soft Graviton Theorem in Generic Theories of Quantum Gravity,''
  arXiv:1706.00759 [hep-th].
 
 \bibitem{1707.06803} 
  S.~Chakrabarti, S.~P.~Kashyap, B.~Sahoo, A.~Sen and M.~Verma,
  ``Subleading Soft Theorem for Multiple Soft Gravitons,''
  arXiv:1707.06803 [hep-th].


\bibitem{1312.2229} 
  A.~Strominger,
  ``On BMS Invariance of Gravitational Scattering,''
  JHEP {\bf 1407}, 152 (2014)
  doi:10.1007/JHEP07(2014)152
  [arXiv:1312.2229 [hep-th]].

\bibitem{1401.7026} 
  T.~He, V.~Lysov, P.~Mitra and A.~Strominger,
  ``BMS supertranslations and WeinbergÕs soft graviton theorem,''
  JHEP {\bf 1505}, 151 (2015)
  doi:10.1007/JHEP05(2015)151
  [arXiv:1401.7026 [hep-th]].

\bibitem{1408.2228} 
  M.~Campiglia and A.~Laddha,
  ``Asymptotic symmetries and subleading soft graviton theorem,''
  Phys.\ Rev.\ D {\bf 90}, no. 12, 124028 (2014)
  doi:10.1103/PhysRevD.90.124028
  [arXiv:1408.2228 [hep-th]].

\bibitem{1411.5745} 
  A.~Strominger and A.~Zhiboedov,
  ``Gravitational Memory, BMS Supertranslations and Soft Theorems,''
  JHEP {\bf 1601}, 086 (2016)
  doi:10.1007/JHEP01(2016)086
  [arXiv:1411.5745 [hep-th]].

\bibitem{1502.02318} 
  M.~Campiglia and A.~Laddha,
  ``New symmetries for the Gravitational S-matrix,''
  JHEP {\bf 1504}, 076 (2015)
  doi:10.1007/JHEP04(2015)076
  [arXiv:1502.02318 [hep-th]].

\bibitem{1502.06120} 
  S.~Pasterski, A.~Strominger and A.~Zhiboedov,
  ``New Gravitational Memories,''
  JHEP {\bf 1612}, 053 (2016)
  doi:10.1007/JHEP12(2016)053
  [arXiv:1502.06120 [hep-th]].

\bibitem{1502.07644} 
  D.~Kapec, V.~Lysov, S.~Pasterski and A.~Strominger,
  ``Higher-Dimensional Supertranslations and Weinberg's Soft Graviton Theorem,''
  Annals of Mathematical Sciences and Applications, Volume 2 (2017),
  pp 69-94
  doi:10.4310/AMSA.2017.v2.n1.a2
  [arXiv:1502.07644 [gr-qc]].
 
 \bibitem{1505.05346} 
  M.~Campiglia and A.~Laddha,
  ``Asymptotic symmetries of QED and Weinberg?s soft photon theorem,''
  JHEP {\bf 1507}, 115 (2015)
  doi:10.1007/JHEP07(2015)115
  [arXiv:1505.05346 [hep-th]].
  
\bibitem{1506.05789} 
  S.~G.~Avery and B.~U.~W.~Schwab,
  ``Burg-Metzner-Sachs symmetry, string theory, and soft theorems,''
  Phys.\ Rev.\ D {\bf 93}, 026003 (2016)
  doi:10.1103/PhysRevD.93.026003
  [arXiv:1506.05789 [hep-th]].

\bibitem{1509.01406} 
  M.~Campiglia and A.~Laddha,
  ``Asymptotic symmetries of gravity and soft theorems for massive particles,''
  JHEP {\bf 1512}, 094 (2015)
  doi:10.1007/JHEP12(2015)094
  [arXiv:1509.01406 [hep-th]].

\bibitem{1605.09094} 
  M.~Campiglia and A.~Laddha,
  ``Sub-subleading soft gravitons: New symmetries of quantum gravity?,''
  Phys.\ Lett.\ B {\bf 764}, 218 (2017)
  doi:10.1016/j.physletb.2016.11.046
  [arXiv:1605.09094 [gr-qc]].

\bibitem{1605.09677} 
  M.~Campiglia and A.~Laddha,
  ``Subleading soft photons and large gauge transformations,''
  JHEP {\bf 1611}, 012 (2016)
  doi:10.1007/JHEP11(2016)012
  [arXiv:1605.09677 [hep-th]].

\bibitem{1608.00685} 
  M.~Campiglia and A.~Laddha,
  ``Sub-subleading soft gravitons and large diffeomorphisms,''
  JHEP {\bf 1701}, 036 (2017)
  doi:10.1007/JHEP01(2017)036
  [arXiv:1608.00685 [gr-qc]].

\bibitem{1612.08294} 
  E.~Conde and P.~Mao,
  ``BMS Supertranslations and Not So Soft Gravitons,''
  arXiv:1612.08294 [hep-th].

\bibitem{1701.00496} 
  T.~He, D.~Kapec, A.~M.~Raclariu and A.~Strominger,
  ``Loop-Corrected Virasoro Symmetry of 4D Quantum Gravity,''
  arXiv:1701.00496 [hep-th].


\bibitem{1612.05886} 
  M.~Asorey, A.~P.~Balachandran, F.~Lizzi and G.~Marmo,
  ``Equations of Motion as Constraints: Superselection Rules, Ward Identities,''
  arXiv:1612.05886 [hep-th].

\bibitem{1703.01351} 
  A.~Campoleoni, D.~Francia and C.~Heissenberg,
  ``On higher-spin supertranslations and superrotations,''
  JHEP {\bf 1705}, 120 (2017)
  doi:10.1007/JHEP05(2017)120
  [arXiv:1703.01351 [hep-th]].

\bibitem{1703.05448} 
  A.~Strominger,
  ``Lectures on the Infrared Structure of Gravity and Gauge Theory,''
  arXiv:1703.05448 [hep-th].

\bibitem{1707.08016} 
  M.~Pate, A.~M.~Raclariu and A.~Strominger,
  ``Color Memory,''
  arXiv:1707.08016 [hep-th].

\bibitem{1709.03850} 
  A.~Laddha and P.~Mitra,
  ``Asymptotic Symmetries and Subleading Soft Photon Theorem in Effective Field Theories,''
  arXiv:1709.03850 [hep-th].
  
\bibitem{1711.04371} 
  D.~Kapec and P.~Mitra,
  ``A $d$-Dimensional Stress Tensor for Mink$_{d+2}$ Gravity,''
  arXiv:1711.04371 [hep-th].

\bibitem{1712.01204} 
  M.~Pate, A.~M.~Raclariu and A.~Strominger,
  ``Gravitational Memory in Higher Dimensions,''
  arXiv:1712.01204 [hep-th].

\bibitem{1712.09591} 
  A.~Campoleoni, D.~Francia and C.~Heissenberg,
  ``Asymptotic Charges at Null Infinity in Any Dimension,''
  Universe {\bf 4}, no. 3, 47 (2018)
  doi:10.3390/universe4030047
  [arXiv:1712.09591 [hep-th]].

\bibitem{1803.03023} 
  A.~H.~Anupam, A.~Kundu and K.~Ray,
  ``Double Soft Graviton Theorems and BMS Symmetries,''
  arXiv:1803.03023 [hep-th].

\bibitem{1801.07719} 
  A.~Laddha and A.~Sen,
  ``Gravity Waves from Soft Theorem in General Dimensions,''
  arXiv:1801.07719 [hep-th].

\bibitem{0912.4254} 
  W.~D.~Goldberger and A.~Ross,
  ``Gravitational radiative corrections from effective field theory,''
  Phys.\ Rev.\ D {\bf 81}, 124015 (2010)
  doi:10.1103/PhysRevD.81.124015
  [arXiv:0912.4254 [gr-qc]].
  
\bibitem{1203.2962} 
  R.~A.~Porto, A.~Ross and I.~Z.~Rothstein,
  ``Spin induced multipole moments for the gravitational wave amplitude from binary inspirals to 2.5 
  Post-Newtonian order,''
  JCAP {\bf 1209}, 028 (2012)
  doi:10.1088/1475-7516/2012/09/028
  [arXiv:1203.2962 [gr-qc]].

\bibitem{1211.6095} 
  W.~D.~Goldberger, A.~Ross and I.~Z.~Rothstein,
  ``Black hole mass dynamics and renormalization group evolution,''
  Phys.\ Rev.\ D {\bf 89}, no. 12, 124033 (2014)
  doi:10.1103/PhysRevD.89.124033
  [arXiv:1211.6095 [hep-th]].

\bibitem{bloch} 
  F.~Bloch and A.~Nordsieck,
  ``Note on the Radiation Field of the electron,''
  Phys.\ Rev.\  {\bf 52}, 54 (1937).
  doi:10.1103/PhysRev.52.54

\bibitem{kinoshita} 
  T.~Kinoshita,
  ``Mass singularities of Feynman amplitudes,''
  J.\ Math.\ Phys.\  {\bf 3}, 650 (1962).
  doi:10.1063/1.1724268

\bibitem{lee} 
  T.~D.~Lee and M.~Nauenberg,
  ``Degenerate Systems and Mass Singularities,''
  Phys.\ Rev.\  {\bf 133}, B1549 (1964).
  doi:10.1103/PhysRev.133.B1549

\bibitem{kulish}
P.~P.~Kulish and L.~D.~Faddeev, ÒAsymptotic conditions and infrared divergences 
in quantum electrodynamics,Ó Theor. Math. Phys. {\bf 4}, 745 (1970) 
[Teor. Mat. Fiz. {\bf 4}, 153 (1970)].
  
\bibitem{1308.6285} 
  J.~Ware, R.~Saotome and R.~Akhoury,
  ``Construction of an asymptotic S matrix for perturbative quantum gravity,''
  JHEP {\bf 1310}, 159 (2013)
  doi:10.1007/JHEP10(2013)159
  [arXiv:1308.6285 [hep-th]].

\bibitem{1705.04311} 
  D.~Kapec, M.~Perry, A.~M.~Raclariu and A.~Strominger,
  ``Infrared Divergences in QED, Revisited,''
  Phys.\ Rev.\ D {\bf 96}, no. 8, 085002 (2017)
  doi:10.1103/PhysRevD.96.085002
  [arXiv:1705.04311 [hep-th]].

\bibitem{jackson}
J.D.~Jackson, ``Classical Electrodynamics,''

\bibitem{peter}
P.~C.~Peters, ``Relativistic Gravitational Bremsstrahlung.'', Phys.\ Rev.\ {\bf D1}, 1559 (1970).


\end{thebibliography}
\end{document}